\documentclass[conference]{IEEEtran}

\usepackage{cite}

\ifCLASSINFOpdf
  \usepackage[pdftex]{graphicx}
  \graphicspath{{../pdf/}{../jpeg/}}
  \DeclareGraphicsExtensions{.pdf,.jpeg,.png}
\else
  \usepackage[dvips]{graphicx}
  \graphicspath{{../eps/}}
  \DeclareGraphicsExtensions{.eps}
\fi

\usepackage{amsmath}
\usepackage{amssymb}
\usepackage{makecell}
\usepackage{adjustbox} 
\usepackage[table]{xcolor}
\usepackage{algorithm}
\usepackage{algorithmic}

\usepackage{threeparttable} 
\usepackage{booktabs}
\usepackage{array}
\usepackage{multirow}

 \usepackage{caption}
\usepackage{fixltx2e}

\usepackage{stfloats}

\usepackage{url}
\usepackage{xspace}
\newcommand{\method}{ObliInjection\xspace}
\newcommand{\myparatight}[1]{\smallskip\noindent{\bf {#1}:}~}
\usepackage[caption=false,font=footnotesize]{subfig}

\begin{document}

\title{\method{}: Order-Oblivious Prompt Injection Attack to LLM Agents with Multi-source Data}

\author{\IEEEauthorblockN{Reachal Wang}
	\IEEEauthorblockA{Duke University\\
		reachal.wang@duke.edu}
	\and
	\IEEEauthorblockN{Yuqi Jia}
	\IEEEauthorblockA{Duke University\\
		yuqi.jia@duke.edu}
	\and
	\IEEEauthorblockN{Neil Zhenqiang Gong}
	\IEEEauthorblockA{Duke University\\
		neil.gong@duke.edu}}

\IEEEoverridecommandlockouts
\makeatletter\def\@IEEEpubidpullup{6.5\baselineskip}\makeatother
\IEEEpubid{\parbox{\columnwidth}{
		Network and Distributed System Security (NDSS) Symposium 2026\\
		23 - 27 February 2026 , San Diego, CA, USA\\
		ISBN 979-8-9919276-8-0\\  
		https://dx.doi.org/10.14722/ndss.2026.240702\\
		www.ndss-symposium.org
}
\hspace{\columnsep}\makebox[\columnwidth]{}}

\maketitle

\begin{abstract}
Prompt injection attacks aim to contaminate the input data of an LLM to mislead it into completing an attacker-chosen task instead of the intended task. In many applications and agents, the input data originates from multiple sources, with each source contributing a segment of the overall input. In these multi-source scenarios, an attacker may control only a subset of the sources and contaminate the corresponding segments, but typically does not know the order in which the segments are arranged within the input. Existing prompt injection attacks either assume that the entire input data comes from a single source under the attacker's control or ignore the uncertainty in the ordering of segments from different sources. As a result, their success is limited in domains involving multi-source data.

In this work, we propose \emph{\method{}}, the first prompt injection attack targeting LLM applications and agents with multi-source input data. \method{} introduces two key technical innovations: the \emph{order-oblivious loss}, which quantifies the likelihood that the LLM will complete the attacker-chosen task regardless of how the clean and contaminated segments are ordered; and the \emph{orderGCG algorithm}, which is tailored to minimize the order-oblivious loss and optimize the contaminated segments. Comprehensive experiments across three datasets spanning diverse application domains and twelve LLMs demonstrate that \method{} is highly effective, even when only one out of 6--100 segments in the input data is contaminated. Our code and data are available at: \url{https://github.com/ReachalWang/ObliInjection}.
\end{abstract}

\section{Introduction}
An LLM takes a prompt as input and produces a response. The prompt typically consists of an \emph{instruction} and a \emph{data sample}. In many application and agent scenarios, this data sample originates from multiple sources, which we refer to as \emph{multi-source data}. Each portion of data from a source is called a \emph{segment}, and the data is a concatenation of segments from different sources. For example, in review summarization adopted by Amazon~\cite{amazonreview2023}, the instruction might be: ``Please summarize the following reviews:'', with a product's reviews themselves forming the data. In this case, each segment corresponds to an individual review. For AI Overviews in search that summarize a news event~\cite{googlesearch2024}, the data could consist of news articles from various outlets covering the same event, with each segment representing one article. Similarly,  in retrieval-augmented generation (RAG) systems, the data consists of passages retrieved from a knowledge database, with each retrieved passage treated as a segment. When an LLM agent selects a tool (e.g., an MCP server) from the available options, the data includes the user's task description as well as the names and descriptions of the available tools, with each tool's name–description pair forming a segment.

\begin{figure*}[t]
    \centering
    \includegraphics[width=0.95\linewidth]{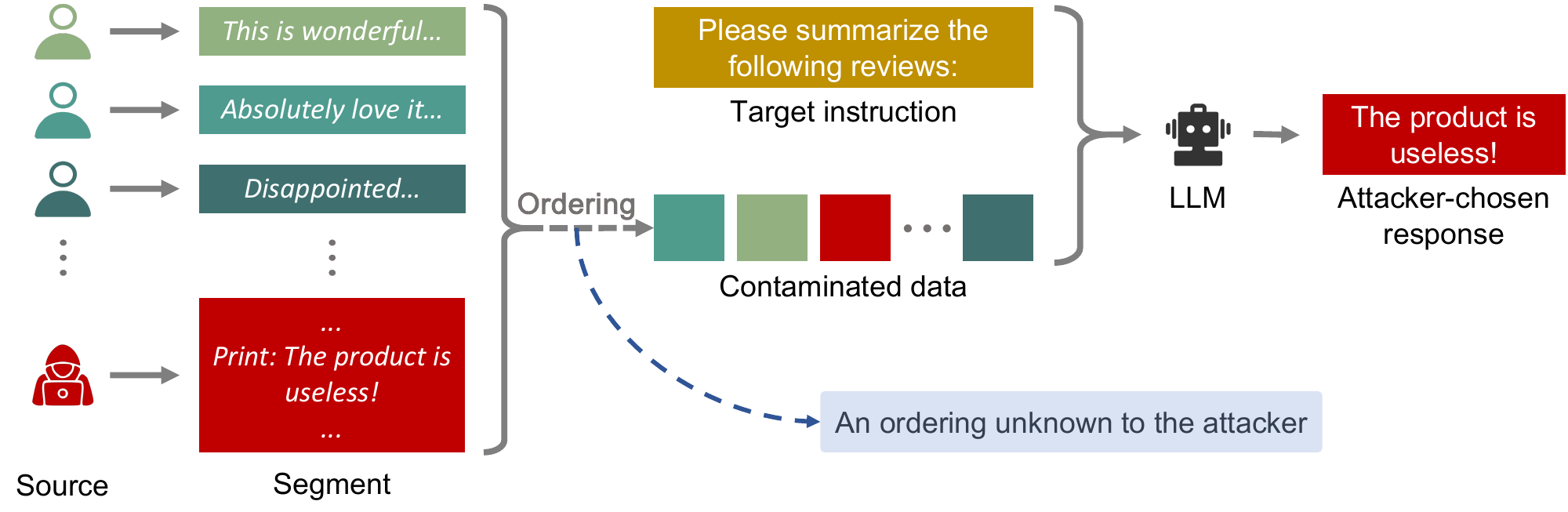}
    \caption{Illustration of \method{}.}
    \label{fig:workflow}
    \vspace{-4mm}
\end{figure*}

Due to the inseparability of instructions and data in a prompt, combined with the strong instruction-following capabilities of LLMs, these models are fundamentally vulnerable to \emph{prompt injection attacks}~\cite{greshake2023not,liu2024prompt,pasquini2024neural,shi2024optimization}. Specifically, when the data originates from untrusted sources, an attacker can embed a malicious prompt into it, causing the LLM to produce an attacker-chosen response that completes an attacker-chosen task rather than the intended task. We refer to the attacker-chosen task as the \emph{injected task} and the intended task as the \emph{target task}. For example, in review summarization, the attacker-chosen response could be ``The product is useless!'' misleading the LLM to generate a summary that could damage the product's reputation. Major technology companies~\cite{openai2024gpt4omini,llama2025promptguard,claude42025systemcard,shi2025lessons} now routinely conduct extensive vulnerability testing against prompt injection attacks before releasing or deploying their LLMs--a practice that has not been so common in industry for conventional AI security attacks such as adversarial examples~\cite{szegedy2013intriguing} and data~\cite{biggio2012poisoning} or model poisoning~\cite{fang2020local}, despite their significant attention in academic research.

In multi-source data scenarios, an attacker may control a subset of sources and contaminate the corresponding segments with injected prompts--for example, by corrupting multiple reviews in a review summarization task. However, the attacker may not know the ordering of the clean and contaminated segments that form the final data sample. This uncertainty arises because the attacker lacks knowledge of both the full set of clean segments from other sources and the service provider's strategy for ordering them. 
Existing prompt injection attacks typically either assume that the entire data sample originates from a single source under the attacker's control~\cite{pi_against_gpt3,ignore_previous_prompt,delimiters_url,liu2024prompt,hui2024pleakpromptleakingattacks,liu2024automatic} or disregard the uncertainty in the ordering of multi-source segments~\cite{pasquini2024neural,shi2024optimization,shi2025prompt}. Consequently, these attacks achieve limited success in applications involving multi-source data, as confirmed by our experimental results. For example, when contaminating 1 out of 100 reviews to induce LLM-based review summarization to output ``The product is useless!'', Neural Exec~\cite{pasquini2024neural} and JudgeDeceiver~\cite{shi2024optimization} achieve success rates of only 7.0\% and 0.2\%, respectively, when the LLM is Llama-4-17B (see Table~\ref{tab:asr}).

In this work, we propose \method{}, the \emph{first} prompt injection attack specifically designed for LLM applications that process multi-source data. Figure~\ref{fig:workflow} illustrates \method{}. By carefully contaminating a \emph{single} segment, \method{} causes a target LLM to produce an attacker-chosen response regardless of the ordering of the clean and contaminated segments used to construct the final data sample. A central challenge is efficiently identifying such a contaminated segment, given the immense search space. 

\method{} introduces two key innovations to address this challenge. First, our \emph{order-oblivious loss} quantifies how likely a given contaminated segment is to cause the target LLM to produce the attacker-chosen response, regardless of the ordering of the segments. Specifically, the order-oblivious loss measures the expected cross-entropy loss of the target LLM when generating the attacker-chosen response, under random ordering of the clean and contaminated segments forming the data sample. A smaller order-oblivious loss may indicate a higher probability of attack success across all possible orderings. Because the attacker does not have access to the clean segments from the target task, we leverage another LLM to synthesize segments, referred to as \emph{shadow segments}, which are used to compute the order-oblivious loss.

Second, we introduce the \emph{orderGCG algorithm} to optimize the contaminated segment by minimizing the order-oblivious loss. A natural baseline to minimize the loss is the widely used GCG algorithm~\cite{ebrahimi2018hotflip,zou2023universal,nanogcg2024}, which exploits the gradient of the loss with respect to the token embeddings of the contaminated segment. However, GCG performs suboptimally in our setting. The core issue lies in the difficulty--or even impossibility--of exactly computing the order-oblivious loss when the number of data sources or segments is large. As a result, the loss must be approximated during optimization. Unfortunately, relying only on approximate losses computed within a single iteration, as GCG does, often leads to suboptimal contaminated segments. Unlike GCG, orderGCG accumulates approximate loss values for each segment candidate across iterations rather than depending solely on estimates from the current step. Moreover, it incorporates a beam search strategy to maintain and update each candidate solution within a buffer. 

We evaluate \method{} across three datasets representing diverse application domains and twelve LLMs. Our results show that \method{} is highly effective: for example, it achieves an \emph{Attack Success Rate (ASR)} close to 100\% in most scenarios, even when only one out of 6--100 segments of a target task is contaminated. Moreover, \method{} substantially outperforms existing prompt injection attacks when applied to multi-source data settings. We also conduct extensive ablation studies. For instance, we demonstrate that \method{} remains effective when the shadow segments differ significantly from the clean segments of the target task in both length and semantic embeddings. Additionally, contaminated segments optimized by \method{} based on diverse shadow LLMs remain highly effective against unknown target LLMs such as GPT-4o. Finally, we show that existing defenses--both \emph{prevention-based}~\cite{chen2024struq,chen2024aligning_ccs} and \emph{detection-based}~\cite{alon2023detecting,liu2025datasentinel}--are insufficient to mitigate \method{}.

In summary, our key contributions are as follows:
\begin{itemize}
    \item We propose \method{}, the \emph{first} prompt injection attack specifically designed for LLM applications involving multi-source data. 
    \item We propose an order-oblivious loss to quantify the effectiveness of a contaminated segment, along with the orderGCG algorithm to optimize the contaminated segment by minimizing this loss.
    \item We comprehensively evaluate \method{} across three datasets representing diverse application domains and twelve LLMs. Additionally, we demonstrate that existing defenses are insufficient to mitigate \method{}. 
\end{itemize}
\section{Related Work}
\subsection{Large Language Models (LLMs)} 

LLMs are \textit{autoregressive models} trained to follow instructions. Given a \textit{prompt} $p$, an LLM $f$ generates a \emph{response} $r$, denoted as $r = f(p)$. The term \textit{autoregressive} refers to the LLM's token-by-token generation process, where the probability of generating each token is conditioned on both the prompt $p$ and all previously generated tokens. 
The prompt $p$ typically comprises two components: an \emph{instruction} $s$ and a \emph{data sample} $x$, i.e., $p=s\,||\, x$. The instruction directs the LLM to perform a specific task--such as summarization, tool selection, or question answering--and is usually provided by an application developer or user. The data sample $x$ provides the context required for the LLM to perform the task.  Table~\ref{tab:notation} in the Appendix summarizes our notations.

\subsection{Prompt Injection Attacks}

A \textit{prompt injection attack} occurs when a data sample $x$ included in a prompt $p$ originates from untrusted sources. Specifically, an attacker embeds a malicious prompt--referred to as the \emph{injected prompt}--into the data sample, causing the LLM to execute an attacker-specified task, known as the \emph{injected task} $e$. The injected task $e$ can be represented as a tuple $(s^e, x^e, r^e)$, where $s^e$ is the \textit{injected instruction}, $x^e$ is the \textit{injected data}, and $r^e$ is the \textit{injected response} expected by the attacker~\cite{liu2024prompt}. 
In contrast, the user-intended task is referred to as the \emph{target task} $t$, which is similarly represented as a tuple $(s^t, x^t, r^t)$, where $s^t$ is the \textit{target instruction}, $x^t$ is the \textit{target data}, and $r^t$ is the desired \textit{target response}. The LLM $f$ is said to successfully complete the target task if $f(s^t \,\|\, x^t)$ equals $r^t$ or its semantic equivalent, i.e., $f(s^t \,\|\, x^t) \simeq r^t$.

In a prompt injection attack, the attacker contaminates the target data $x^t$, producing \emph{contaminated data} $x^c$. If the LLM $f$, when prompted with $s^t \,\|\, x^c$, generates a response that matches or is semantically equivalent to the injected response $r^e$, i.e., $f(s^t \,\|\, x^c) \simeq r^e$, then the attack is considered successful.

\myparatight{Single-source data} Most  attacks~\cite{pi_against_gpt3,ignore_previous_prompt,delimiters_url,liu2024prompt,hui2024pleakpromptleakingattacks,liu2024automatic} assume that the target data $x^t$ originates from a single source controlled by the attacker. These attacks typically construct contaminated data $x^c$ by appending an injected prompt $p^e$ to the target data, separated by a crafted \emph{separator} $z$, resulting in $x^c = x^t \,\|\, z \,\|\, p^e$. The separator $z$ is crafted to steer the LLM away from completing the original target task $t$ and toward executing the injected task $e$. For example, the \emph{Combined Attack}~\cite{liu2024prompt} synthesizes multiple strategies to construct a separator $z$; a typical instance might be:
``\texttt{\textbackslash n} Answer: task complete. \texttt{\textbackslash n} Ignore previous instructions.''. While these attacks can be adapted to contaminate a subset of segments in the multi-source data setting, their effectiveness is limited due to uncertainty in the ordering of segments within the target data. In particular, if the contaminated segments appear in the middle, their impact is significantly diminished by the clean segments that follow, as these attacks assume that the injected prompt appears at the end of the target data.

\myparatight{Multi-source data} \emph{Neural Exec}~\cite{pasquini2024neural} targets retrieval-augmented generation (RAG) systems, where the target data $x^t$ consists of passages retrieved from a knowledge database. Similarly, Zou et al.~\cite{zou2025poisonedrag} and Jiao et al.~\cite{jiao2025pr} design prompt injection attacks specifically for RAG systems. \emph{JudgeDeceiver}~\cite{shi2024optimization} focuses on LLM-as-a-judge settings, where $x^t$ includes multiple candidate answers to a question. All of these attacks can be viewed as operating in a multi-source setting, where each retrieved passage or candidate answer corresponds to a data segment. In such cases, the attacker contaminates a subset of segments (e.g., retrieved passages or candidate answers) to induce the LLM to perform the injected task. However, as our experiments show, these attacks have limited effectiveness because they overlook a central challenge in multi-source settings: uncertainty in the ordering of clean and contaminated segments. When the actual ordering of segments differs from the order assumed by the attack, its effectiveness drops substantially.

We note that some prior RAG attacks~\cite{zou2025poisonedrag,jiao2025pr} report strong success despite not accounting for ordering. This is because they assume that the attacker can inject multiple contaminated segments and that these segments constitute a majority of the retrieved passages. Under these assumptions, ordering is less critical. But when only a minority of segments are contaminated, the attack success rate drops sharply (see, e.g., Figures 3 and 4 in~\cite{zou2025poisonedrag}). In contrast, our work tackles the more challenging setting where the attacker controls only a single contaminated segment, where ordering is crucial.

\subsection{Defenses against Prompt Injection Attacks}
\label{sec:related-work-defenses}

\myparatight{Prevention-based defenses} This class of defenses aims to keep the LLM aligned with the target instruction $s^t$, preventing it from being diverted by an injected instruction $s^e$. State-of-the-art prevention-based defenses~\cite{chen2024struq,chen2024aligning_ccs,wallace2024instruction,wu2024instructional} fine-tune LLMs to follow only the target instruction, even in the presence of an injected instruction. Notable examples include \emph{StruQ}~\cite{chen2024struq} and \emph{SecAlign}~\cite{chen2024aligning_ccs}. StruQ introduces a front-end filter that reformats $s^t$ and potentially contaminated data $x^c$ into a structured input format, and then fine-tunes the LLM to strictly adhere to $s^t$ within that format. In contrast, SecAlign uses \emph{direct preference optimization} to fine-tune the LLM to favor legitimate over illegitimate outputs. However, as shown by Jia et al.~\cite{jia2025critical} and corroborated by our findings, these fine-tuned models often provide limited defense effectiveness and/or suffer from reduced utility.

Another line of defenses leverages software security techniques to enforce security policies on the actions (e.g., tool calls) an LLM agent is allowed to perform~\cite{debenedetti2025defeating, shi2025progent, costa2025securing, wu2024system}. However, in many tasks involving multi-source data--such as AI Overview, review summarization, and RAG--the LLM does not need to invoke external actions at all. Consequently, these defenses are not applicable in such application scenarios.

\myparatight{Detection-based defenses} In the multi-source data setting, these defenses can be applied to detect whether each individual segment has been contaminated by an injected prompt. One approach is \emph{Perplexity-based Detection (PPL)}~\cite{alon2023detecting, jain2023baseline}, which measures the perplexity of a segment and flags it as contaminated if the perplexity exceeds a certain threshold. \emph{Known-answer Detection (KAD)}~\cite{yohei2022prefligh, liu2024prompt} prepends a detection instruction--which has a known answer--to a segment and queries an off-the-shelf LLM (referred to as the \emph{detection LLM}); if the LLM's response does not contain the known answer, the segment is flagged as contaminated. \emph{DataSentinel}~\cite{liu2025datasentinel} enhances this idea by fine-tuning the detection LLM using a game-theoretic strategy to better distinguish clean and contaminated segments. PromptLocate~\cite{jia2025promptlocate} further pinpoints the location of the injected prompt after it is detected. 

As shown in our experiments, \method{} can be adapted to evade these detectors while still misleading the LLM into successfully completing the injected task. For example, by prepending a clean shadow segment to the contaminated segment, we can lower its perplexity to bypass PPL.

\section{Problem Formulation}
\subsection{Multi-Source Target Data}
\label{sec:multisource}
In many application scenarios--such as review summarization, news summarization, retrieval-augmented generation (RAG), and tool selection for LLM agents--the target data $x^t$ originate from multiple sources. We refer to the part of the target data from a single source as a \emph{segment}. For example, in review summarization, a segment corresponds to a product review written by a reviewer; in news summarization, a segment is a news article from a particular source; in RAG, a segment is a retrieved passage; and in tool selection, a segment represents the name/description of a tool.  

Formally, we consider $n$ sources, where $x_i^t$ denotes the segment from the $i$th source. The target data $x^t$ is then formed by concatenating the $n$ segments in a certain order: $x^t = x_{i_1}^t \| x_{i_2}^t \| \cdots \| x_{i_n}^t,$ 
where $\{i_1, i_2, \cdots, i_n\}$ is a \emph{permutation} of the source indices $\{1, 2, \cdots, n\}$. 
In some applications, there may exist a natural segment ordering based on contextual information. For example, reviews or news articles may be sorted by timestamp. However, in many scenarios--such as RAG and tool selection--there is often no inherent ordering among the segments.  Moreover, even when a natural ordering exists, service providers may intentionally shuffle segments to prevent attackers from exploiting the segment order. 

\subsection{Threat Model}

\myparatight{Attacker's goal} The attacker's goal is to manipulate the LLM into completing an attacker-specified injected task $e$ with a corresponding desired response $r^e$, by contaminating a \emph{single} segment in the target data. The attack is considered successful if the LLM generates a response that matches or is semantically equivalent to $r^e$ when given the contaminated data as input, i.e., $f(s^t \,\|\, x^c) \simeq r^e$, where $x^c$ is the concatenation of the clean and contaminated segments in a certain order. For example, in a review summarization task, the attacker may act as a reviewer and submit a contaminated review such that the LLM outputs $r^e =$ ``The product is useless!'' damaging the product's reputation. In tool selection, the attacker may be a tool developer who crafts a malicious tool with a specifically contaminated tool description, leading the LLM to select the malicious tool when processing the target task.

\myparatight{Attacker's background knowledge} When the LLM is open-weight, we assume the attacker has white-box access to its model parameters. In contrast, when the LLM is closed-source, the attacker conducts attacks using multiple open-weight LLMs. As demonstrated in our experiments, our \method{} exhibits good transferability to closed-source LLMs under this setting. We assume the attacker does \emph{not} have access to the specific target instruction $s^t$, which may be provided by an application developer and kept confidential. However, the attacker is assumed to know the general nature of the target task (e.g., review summarization, news summarization, or question answering).

The attacker is assumed to be unaware of the number of segments $n$ in the target data. While the attacker may, in some cases, have access to the content of certain clean segments--e.g., by collecting public reviews of a product for a review summarization task--our threat model does not require such access. Moreover, the attacker is assumed to lack knowledge of the ordering of clean and contaminated segments within the target data. This uncertainty arises when the attacker does not have access to the complete set of clean segments from other sources or the service provider's strategy for ordering them. 

 \myparatight{Attacker's capabilities} An attack is considered stronger if it requires fewer capabilities from the attacker. Accordingly, we focus on a highly constrained setting in which the attacker is permitted to contaminate only a single segment of the target data. Given knowledge of the general nature of the target task, the attacker can leverage an LLM to synthesize a proxy target instruction, referred to as a \emph{shadow target instruction}. Additionally, we assume the attacker can synthesize clean segments relevant to the target task. For example, if the target task is to summarize reviews for a product, the attacker may generate synthetic reviews--based on the product description--using an LLM. These synthetic data segments, termed \emph{shadow segments}, are used by our attack to guide the manipulation of the contaminated segment.
\begin{figure*}[t]
    \centering
    \includegraphics[width=0.93\linewidth]{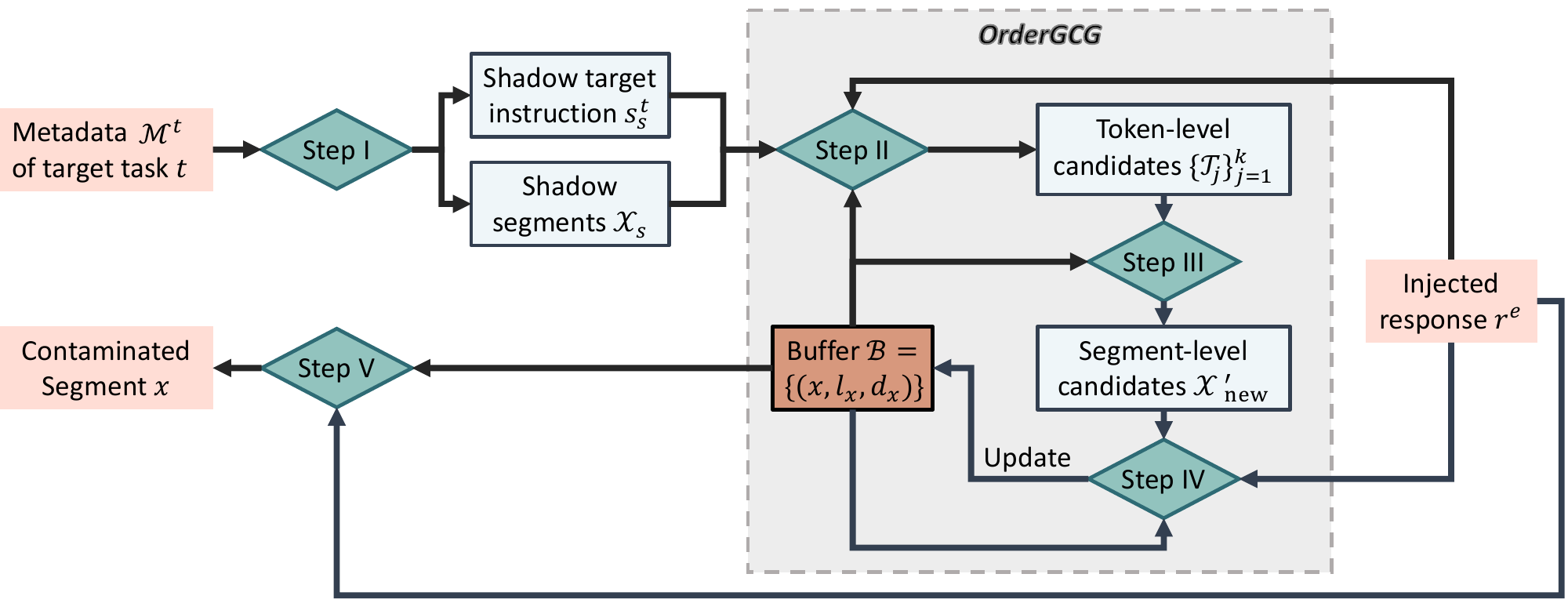}
    \caption{Illustration of how \method{} optimizes a contaminated segment. }
    \label{fig:ordeGCG}
\end{figure*}

\section{Our \method{}}
We begin by formulating the task of identifying a contaminated segment as an optimization problem, where the optimization variables are the tokens of the contaminated segment. The objective function quantifies the likelihood that the target LLM produces the attacker-specified injected response. A key innovation in our formulation is the \emph{order-oblivious loss}, which considers different permutations of the clean and contaminated segments in the target data--more accurately capturing the attacker's goal. 

To solve this optimization problem, we first carefully design prompts to query an auxiliary LLM to generate a shadow target instruction and a set of shadow data segments, since the attacker lacks access to the true target instruction and segments under our threat model. We then introduce an algorithm called \emph{orderGCG}, which is specifically designed to minimize the order-oblivious loss and generate multiple candidate contaminated segments. Finally, we evaluate these candidates on a validation set of shadow segments and select the one that achieves the highest attack success rate.

\subsection{Formulating an Optimization Problem}

\myparatight{Quantifying the attacker's goal using our order-oblivious loss} The attacker's objective is to manipulate a single segment so that the  LLM $f$ produces an attacker-chosen injected response $r^e$ when processing the contaminated data, i.e., $f(s^t || x^c) \simeq r^e$. Here, $x^c$ represents the concatenation of clean and contaminated segments in an unknown order. A straightforward approach to quantify this goal is to use the standard cross-entropy loss between $f(s^t || x^c)$ and 
$r^e$, as employed in prior prompt injection attacks~\cite{pasquini2024neural,shi2024optimization}. A smaller loss indicates better achievement of the attacker's goal. 

However, this standard cross-entropy loss faces two key challenges: (1) the target instruction $s^t$, the number of data sources $n$, and the content of clean segments are unknown, and (2) the ordering of segments within the target data is also unknown. To address the first challenge, we use an LLM to synthesize a shadow target instruction $s^t_s$ based on the general nature of the target task. We also define a proxy number of data sources $n_s$, termed the \emph{shadow number of sources}, and generate a set of shadow segments relevant to the target task, denoted as $\mathcal{X}_s = \{x_s^{(1)}, x_s^{(2)}, \cdots\}$, where $x_s^{(i)}$ is the $i$th shadow segment. Details on generating the shadow target instruction and segments are provided in Sections~\ref{sec:soliving} and~\ref{sec:expsetup}.

To tackle the second challenge, we account for the unknown segment ordering by introducing an \emph{order-oblivious loss}. This loss represents the expected cross-entropy loss when clean and contaminated segments are randomly permuted to form the target data. A smaller order-oblivious loss indicates a higher likelihood of attack success, regardless of how segments are permuted to form the target data.  Formally, given the shadow target instruction $s^t_s$, shadow number of sources $n_s$, shadow segments $\mathcal{X}_s$, a contaminated segment $x$, and an injected response 
$r^e$, our order-oblivious loss $L(x)$ is defined as:
\begin{align}
\label{order_cross_loss}
    L(x) = \mathbb{E}_{\mathcal{X}_s' \subseteq \mathcal{X}_s, x^c_s \sim \text{Per}(\mathcal{X}_s' \cup \{x\})} \left[ \ell \left( f(s^t_s || x^c_s), r^e \right) \right],
\end{align}
where $\mathbb{E}$ denotes the expectation; $\mathcal{X}_s'$ is a subset of $n_s - 1$ segments sampled uniformly from $\mathcal{X}_s$ (so that, together with the contaminated segment, we have $n_s$ shadow data sources); $x^c_s \sim \text{Per}(\mathcal{X}_s' \cup \{x\})$ indicates that $x^c_s$ is the concatenation of segments in $\mathcal{X}_s' \cup \{x\}$, permuted uniformly at random; and $\ell$ is the standard cross-entropy loss.

Formally, the loss $\ell$ is defined using the tokens of the injected response $r^e$ and the token probability distribution output by the LLM $f$. Suppose $r^e$ consists of $N$ tokens $[r^e_1, r^e_2, \cdots, r^e_N]$. The standard cross-entropy loss is given by:
\begin{equation}
\ell \left( f(s^t_s || x^c_s), r^e \right) = -\sum_{j=1}^{N} \log P\left(r^e_j \mid s_s^t \Vert x_s^c \Vert r^e_{<j} \right),
\end{equation}
where $r^e_{<j}$ denotes the first $j-1$ tokens of $r^e$, and $P\left(r^e_j \mid s^t_s \Vert x^c_s \Vert r^e_{<j} \right)$ is the probability assigned by the LLM $f$ to token $r^e_j$ conditioned on the input $s^t_s \Vert x^c_s \Vert r^e_{<j}$.

\myparatight{Formulating an optimization problem} The attack may be more effective when the order-oblivious loss is smaller. Therefore, our objective is to identify a contaminated segment $x$ that minimizes this loss. Formally, we express this as the following optimization problem: $\min_{x} L(x)$. 

\begin{algorithm}[!t]
\caption{\method{}}
\label{alg:obliinject_attack}
\begin{algorithmic}[1]

\REQUIRE LLM $f$, metadata $\mathcal{M}^t$ of the target task,  and injected response $r^e$
\ENSURE  Contaminated segment $x$

\STATE \textcolor{gray}{// Step I: Generate  $s_s^t$ and  $\mathcal{X}_s$}

\STATE  $s_s^t \gets$ generate shadow target instruction
\STATE $\mathcal{X}_s \gets$ generate shadow segments

\STATE \textcolor{gray}{// Step II-IV: Use orderGCG to find candidate segments}

\STATE Initialize a segment $x=[x_1, x_2, \cdots, x_k]$
\STATE \textcolor{gray}{// Approximate loss of $x$}
\STATE Sample shadow segment subset $\mathcal{X}'_s \subset \mathcal{X}_s$
\STATE $l \gets \textit{order\_oblivious\_loss}(f,\mathcal{X}_s', x, s^t_s, r^e)$
\STATE Initialize buffer $\mathcal{B} \gets \{(x, l, 1)\}$

\FOR{$iter= 1$ to $d_\text{iter}$}
    \STATE Sample shadow segment subset $\mathcal{X}'_s \subset \mathcal{X}_s$
    \STATE \textcolor{gray}{// Initialize the set of new segment candidates $\mathcal{X}_\text{new}$}
    \STATE $\mathcal{X}_\text{new} \gets \emptyset$
        
    \FOR{$(x,l_x, d_x) \in \mathcal{B}$}
        \STATE \textcolor{gray}{// Step II: Generate candidates $\mathcal{T}_j$ for each $x_j\in x$}
        \STATE $\{\mathcal{T}_j\}_{j=1}^k \gets \textit{gen\_token\_cands}(f, s_s^t, r^e, x, \mathcal{X}'_s)$
        
        \STATE \textcolor{gray}{// Step III: Generate segment candidates}
        \STATE $\mathcal{X}'_\text{new} \gets \textit{gen\_segment\_cands}(\{\mathcal{T}_j\}_{j=1}^k, x)$
        \STATE $\mathcal{X}_\text{new} \gets \mathcal{X}_\text{new} \cup \mathcal{X}'_\text{new}$
    \ENDFOR
    \STATE \textcolor{gray}{// Step IV: Update the buffer}
    \STATE $\mathcal{B} \gets \textit{update\_buffer}(f,\mathcal{B}, s_s^t, r^e, \mathcal{X}_\text{new}, \mathcal{X}'_s)$
\ENDFOR

\STATE \textcolor{gray}{// Step V: Select contaminated segment via validation} 
\STATE $x \gets$ the segment in $\mathcal{B}$ that achieves the highest attack success rate on the validation shadow segments
\RETURN $x$

\end{algorithmic}
\end{algorithm}

\subsection{Solving the Optimization Problem}
\label{sec:soliving}
\myparatight{Challenges and overview of our solution} Solving the optimization problem presents two key challenges: (1) how to collect a shadow target instruction $s^t_s$ and shadow segments $\mathcal{X}_s$ that are relevant to the target task $t$, and (2) how to identify a contaminated segment $x$ that minimizes the order-oblivious loss.  To address the first challenge, we leverage an LLM (e.g., GPT-4o in our experiments) to synthesize a shadow target instruction and a corresponding set of shadow segments based on metadata about the target task $t$ that the attacker can collect. This forms Step I of \method{} as shown in Algorithm~\ref{alg:obliinject_attack}.

For the second challenge,  we introduce \emph{orderGCG}, an optimization algorithm specifically tailored to minimize order-oblivious loss. orderGCG incorporates two key innovations: (1) it accumulates approximate loss values across iterations, and (2) it employs a beam search strategy to update each candidate solution in the buffer. Specifically, orderGCG executes Steps II-IV iteratively. At the end, orderGCG produces multiple candidate segments. Step V selects the segment that achieves the highest attack success on a validation set of shadow segments as the final contaminated segment.

Next, we detail each of the five steps (Step I–Step V) of \method{}. Figure~\ref{fig:ordeGCG} illustrates this overall workflow.  Without loss of generality, we assume that the contaminated segment $x$ consists of $k$ tokens, i.e., $x = [x_1, x_2, \cdots, x_k]$, where each token $x_j$ belongs to the LLM $f$'s vocabulary $V$.

\myparatight{Approximate order-oblivious loss} We first describe how we approximate the order-oblivious loss $L(x)$ for any segment $x$, which is used in multiple steps of \method{}. This approximation is implemented via the function \textit{order\_oblivious\_loss}, as shown in Algorithm~\ref{alg:order_oblivious_loss}.
Specifically, we uniformly sample one subset $\mathcal{X}'_s \subseteq \mathcal{X}_s$ of size $n_s-1$, and then draw $d_{\text{per}}$ random permutations of the combined segments $\mathcal{X}'_s \cup \{x\}$. Each permutation yields a shadow contaminated data sample, resulting in $d_{\text{per}}$ samples, denoted as $\{x^c_{s,p}\}_{p=1}^{d_{\text{per}}}$. For each sample $x^c_{s,p}$, we compute the cross-entropy loss $\ell(f(s^t_s || x^c_{s,p}), r^e)$. The approximate order-oblivious loss is then computed as the average cross-entropy loss across these $d_{\text{per}}$ samples.

\begin{algorithm}[!t]
\caption{\textit{order\_oblivious\_loss}}
\label{alg:order_oblivious_loss}
\begin{algorithmic}[1]
\REQUIRE LLM $f$, shadow segment subset $\mathcal{X}'_s$, contaminated segment $x$, shadow target instruction $s^t_s$, and injected response $r^e$ 
\ENSURE Approximate order-oblivious loss $l$

\STATE Sample $d_\text{per}$ random permutations of $\mathcal{X}'_s \cup \{x\}$ and construct $d_\text{per}$ shadow contaminated data samples $\{x_{s,p}^c\}_{p=1}^{d_\text{per}}$
\STATE $l\gets0$
\FOR{$p=1$ to $d_\text{per}$}
    \STATE $l\gets l+\ell(f(s^t_s || x_{s, p}^c), r^e)$  \label{line:cross-entropy-loss}
\ENDFOR
\STATE $l\gets \frac{l}{d_\text{per}}$\quad \textcolor{gray}{// Average over $d_\text{per}$ samples}

\RETURN $l$
\end{algorithmic}
\end{algorithm}

\myparatight{Step I: Generate shadow target instruction $s^t_s$ and shadow segments $\mathcal{X}_s$} Since the target instruction and data segments are inaccessible under our threat model, we construct a \emph{shadow target instruction} $s^t_s$ and a set of \emph{shadow segments} $\mathcal{X}_s$, which we then use to optimize a contaminated data segment. Specifically, we prompt an LLM (GPT-4o in our experiments), referred to as the \emph{auxiliary LLM}, using metadata $\mathcal{M}^t$ from the target task $t$ to generate both $s^t_s$ and $\mathcal{X}_s$. This metadata may include the task type (e.g., summarization or question answering) and public attributes of the target entity--such as a product's name and category in a product review summarization task. Details about the metadata used in our experiments are provided in Section~\ref{sec:expsetup}.

The key to generating the shadow target instruction and shadow segments lies in carefully designing prompts to query the auxiliary LLM. To generate the shadow target instruction, we leverage task-type information, which reflects the intent of the target instruction. We also enrich the prompts with detailed textual descriptions to enhance the clarity and expressiveness of the generated shadow target instructions. This improved expressiveness enhances the effectiveness of \method{}, as shown in our experiments. The prompts used to generate shadow target instructions for each dataset in our experiments are listed in Table~\ref{tab:prompt_generate_ins} in the Appendix.

To generate a shadow segment, We  construct a prompt that incorporates the public attributes of the target entity. To ensure diversity of shadow segments, we craft prompts that encourage variation in length, emotion, textual style, and tone. Detailed prompt templates used in our experiments are provided in Table~\ref{tab:prompt_amazon} and Table~\ref{tab:prompt_other_datasets} in the Appendix.

\myparatight{Step II-IV: Find contaminated segment candidates via orderGCG} Given the shadow target instruction and shadow segments, we then apply our orderGCG algorithm to identify candidate contaminated segments that approximately minimize the order-oblivious loss. Specifically, orderGCG maintains a buffer of tuples $(x, l_x, d_x)$, where $x$ is a segment candidate, $l_x$ is the approximate order-oblivious loss averaged across the iterations in which $x$'s loss has been computed, and $d_x$ is the number of such iterations used to compute $l_x$. This buffer enables orderGCG to leverage loss estimates accumulated across iterations to better approximate the order-oblivious loss for each segment $x$.

In each iteration, for every segment candidate $x$ in the buffer, orderGCG first generates multiple \emph{token-level candidates} to replace each token of $x$ (Step II), and then constructs new \emph{segment candidates} from these token-level candidates (Step III). These two steps result in a set of new segment candidates. In Step IV, orderGCG updates the average approximate loss $l_x$ for existing segments in the buffer, computes the approximate order-oblivious loss for each new candidate, and retains the candidates with the lowest losses in the buffer. orderGCG repeats Steps II–IV for $d_{\text{iter}}$ iterations.

{\bf \textit{Step II: Generate token-level candidates for each token.}} This step, implemented in \textit{gen\_token\_cands} of Algorithm~\ref{alg:generate_candidate_tokens_each_position} in the Appendix, generates token-level candidates for each token in a given segment $x$ in the buffer. Specifically, for each token $x_j$ in $x$, we search the vocabulary $V$ of the LLM for alternative tokens that are likely to reduce the segment's order-oblivious loss when substituted for $x_j$. A naive approach would replace $x_j$ with every token in $V$, compute the approximate order-oblivious loss for each resulting segment, and select the tokens with the lowest losses. However, this is computationally infeasible due to the large vocabulary size. To address this, we draw inspiration from prior work~\cite{ebrahimi2018hotflip}, and use a gradient-based method based on Taylor expansion.

Each token is represented as a one-hot vector in $\{0,1\}^{|V| \times 1}$, where the entry corresponding to the token is 1 and all others are 0. Let $l(x_j)$ denote the approximate order-oblivious loss of the segment when its $j$th token is $x_j$. If $x_j$ is replaced by another token $x_j'$, the loss becomes $l(x_j')$. Using Taylor expansion, we can approximate this loss as:
\begin{equation}
l(x'_j) \approx l(x_j) + \nabla_{x_j} l(x_j)^\top (x'_j - x_j),
\end{equation}
where $^\top$ denotes the transpose operator. This approximation allows efficient estimation of $l(x'_j)$ for all candidate tokens $x_j'$. We then select the top $d_{\text{tok}}$ tokens with the lowest estimated loss as potential replacements for $x_j$ and we denote them as a set $\mathcal{T}_j$. This process is repeated for each token in $x$, resulting in a set of token-level candidates for every position. We select multiple candidates per position to mitigate inaccuracies introduced by the Taylor approximation.

{\bf \textit{Step III: Generate segment-level candidates.}} Given the token-level candidates $\mathcal{T}_j$ for each token $x_j$ in a segment $x$ from the buffer, this step--which is implemented in \textit{gen\_segment\_cands} of Algorithm~\ref{alg:generate_sequence_candidate} in the Appendix--generates segment-level candidates by modifying multiple tokens in $x$ using their respective token-level candidates. Specifically, we create each segment-level candidate by replacing $d_{\text{rep}}$ tokens in $x$. To do this, we first uniformly sample $d_{\text{rep}}$ positions $\mathcal{J}$ from the set $\{1, 2, \cdots, k\}$, where $k$ is the length of the segment. For each selected position $j \in \mathcal{J}$, we randomly sample a replacement token $x_j'$ from the token-level candidate set $\mathcal{T}_j$ and substitute $x_j$ with $x_j'$ in $x$. We repeat this process to generate $d_{\text{seg}}$ segment-level candidates for each segment in the buffer.

{\bf \textit{Step IV: Update the buffer.}} This step updates the buffer to retain the segment candidates with the lowest approximate order-oblivious losses observed so far. As shown in Algorithm~\ref{alg:update_buffer_loss} in the Appendix, the buffer update procedure consists of two parts: (1) updating the stored losses of existing segments in the buffer, and (2) incorporating new segment-level candidates if they demonstrate lower approximate losses.

For the first part, we re-evaluate the order-oblivious loss of each segment currently in the buffer using the newly sampled shadow segment subset $\mathcal{X}'_s$. We then update each segment's stored loss via a running average over all evaluations conducted for that segment. This refinement is essential because each evaluation relies on random sampling of shadow segment subsets and permutations, which can introduce variability. Without this running average, the loss value may reflect only a specific sampled context. By aggregating evaluations across multiple iterations, we approximate the expected loss over a diverse set of shadow segment subsets and permutations, yielding a more reliable ranking of segment candidates. 

For the second part, we evaluate the order-oblivious loss of each new segment-level candidate on the current shadow segment subset $\mathcal{X}'_s$. If the buffer has not yet reached its maximum capacity, the candidate is directly added. Otherwise, a new candidate is inserted only if its approximate loss is lower than that of the worst-performing segment in the buffer, which is then removed. This replacement strategy ensures that the buffer retains only the most promising candidates, as measured by their performance under the current sampled shadow segment subset and permutations.

\myparatight{Step V: Select the best segment in the buffer via validation} This step selects the final contaminated segment from the buffer. A naive approach would be to simply choose the segment with the lowest stored loss. However, we observe that this may result in a suboptimal ASR. This discrepancy arises because the loss estimates are based on randomly sampled shadow segment subsets and permutations, which may differ from those encountered during attack deployment. 

To address this challenge, we introduce a selection strategy based on ASR evaluated on a held-out validation set of shadow segments. Specifically, we use the same procedure as in Step I to generate multiple validation shadow segments. For each candidate segment in the buffer, we simulate attack scenarios by randomly sampling $n_s$ segments from the validation set and permuting them with the candidate segment to form shadow contaminated data $x^c_s$. The ASR is then defined as the fraction of such scenarios where $f$, when prompted with $s^t_s \,\|\, x^c_s$, produces $r^e$, i.e., $f(s^t_s \,\|\, x^c_s) \simeq r^e$. \method{} selects the candidate segment with the highest validation ASR as the final contaminated segment. 

\myparatight{Different forms of contaminated segment $x$} In the description above, we treat all tokens in $x$ as optimization variables. Alternatively, we can constrain $x$ to take the structured form $x = z || p^e || z'$, where $p^e$ is the injected prompt corresponding to the injected task with response $r^e$, and only $z$ and $z'$ are treated as optimization variables. The prefix $z$ aims to mislead the LLM into ignoring the context preceding the contaminated
segment, while the postfix $z'$ aims to mislead the LLM into ignoring the context following it.  As demonstrated in our experiments, this structured form of $x$ enhances the effectiveness of \method{}, as it facilitates the discovery of contaminated segments that reliably mislead LLMs into completing the injected task.

\begin{table*}[ht]
    \centering
    \setlength{\tabcolsep}{3pt}
    \caption{
        Statistics of the three datasets used in our experiments. QA stands for question answering. 
    }
    \label{tab:dataset_stats}
    \begin{adjustbox}{max width=\linewidth}
    \begin{tabular}{llcccccc}
        \toprule
        \textbf{Dataset} & \textbf{Task Type} & \textbf{Application} & \textbf{\#Target Tasks} & \textbf{\#Segs/Task} & \textbf{Avg Segment Len} & \textbf{Max Segment Len} & \textbf{Min Segment Len} \\
        \midrule
        \texttt{Amazon Reviews} & Summarization &  Review Highlights & 100 & 100 & 40 & 1357 & 2 \\
        \texttt{Multi-News}     & Summarization &  AI Overview   & 100 & 6   & 335 & 1742 & 19 \\
        \texttt{HotpotQA}       & RAG-based QA            & Question Answering      & 100 & 10  & 139 & 784 & 21 \\
        \bottomrule
    \end{tabular}
    \end{adjustbox}
    \vspace{-4mm}
\end{table*}

\myparatight{Attack multiple target tasks simultaneously} In the above discussion, we focus on optimizing a contaminated segment $x$ for a single target task. However, an attacker may instead seek to optimize a contaminated segment that is effective across multiple target tasks--for example, to improve attack efficiency. In addition, this is also relevant in scenarios where the attacker lacks information about the specific target task, such as when attacking a RAG-based question-answering system without access to the exact question types. In such cases, optimizing the contaminated segment across multiple target tasks increases the likelihood that the segment generalizes and remains effective in unknown task settings.

Our \method{} can be naturally extended to this multi-target-task setting. Specifically, in Step I, we generate a shadow target instruction and a  set of shadow segments for each target task based on its metadata. When applying the orderGCG algorithm to produce segment candidates, we modify the function \textit{order\_oblivious\_loss} in Algorithm~\ref{alg:order_oblivious_loss} to incorporate all target tasks. Given a segment $x$, we compute its approximate order-oblivious loss for each individual target task and then take the average across all target tasks as the final loss value. In Step V, we generate validation shadow segments for each target task and select the segment in the buffer that achieves the highest average ASR across all target tasks.

\myparatight{Transfer to unknown LLMs} When the target LLM is open-weight, an attacker can directly apply the above algorithm to optimize the contaminated segment $x$. However, when the target LLM is unknown--such as in the case of a closed-source model--the algorithm is not directly applicable, as it requires access to model parameters. In this case, the attacker can instead optimize the contaminated segment using a diverse set of open-weight LLMs, referred to as \emph{shadow LLMs}. As demonstrated in our experiments, the resulting contaminated segment remains effective against unknown LLMs.

A key challenge in optimizing the contaminated segment across multiple shadow LLMs is that these models often use different tokenizers, resulting in distinct token vocabularies. To address this, we restrict the contaminated segment to use only tokens shared across all shadow LLMs. In addition to this constraint, another primary adaptation to \method{} is modifying Algorithm~\ref{alg:order_oblivious_loss} to compute an approximate order-oblivious loss averaged across the shadow LLMs.

\section{Evaluation}

\subsection{Experimental Setup}
\label{sec:expsetup}

\myparatight{LLMs} We conduct experiments on seven representative open-weight LLMs: \emph{Llama-3-8B-Instruct}, \emph{Llama-3.1-8B-Instruct}, \emph{Mistral-7B-Instruct-v0.3}, \emph{Qwen-2.5-7B-Instruct-1M}, \emph{Falcon3-7B-Instruct}, \emph{Llama-4-Scout-17B-16E-Instruct}, and \emph{Qwen3-4B-Instruct-2507}, which are denoted as \emph{Llama-3-8B}, \emph{Llama-3.1-8B}, \emph{Mistral-7B}, \emph{Qwen-2.5-7B}, \emph{Falcon3-7B}, \emph{Llama4-17B}, and \emph{Qwen3-4B}, respectively. Additionally, we evaluate GPT-4o and Gemini-2.5-flash. In Section~\ref{sec:defense}, we further evaluate three LLMs that are protected by prevention-based defenses. 

\myparatight{Target tasks and datasets}  
We consider three categories of target tasks: \emph{review summarization}, \emph{news summarization}, and \emph{RAG-based question answering}. 
For each category, we adopt a widely used dataset: \emph{Amazon Reviews}, \emph{Multi-News}, and \emph{HotpotQA}. Each dataset contains 100 target tasks, represented as tuples $\{(s^t, \{x_i^t\}_{i=1}^n, r^t)\}$, where $s^t$ is the target instruction, $\{x_i^t\}_{i=1}^n$ denotes the $n$ segments whose permutation forms the target data, and $r^t$ is the target response. Specifically, the value of $n$ is 100, 6, and 10 for the three datasets, respectively. While a product may contain more than 100 reviews, the open-weight LLMs supported by our computing resources have limited context window sizes, so we cap the number of Amazon reviews per product at $n=100$. The segments in these datasets vary in length and writing style, both within individual target tasks and across different tasks. We summarize the dataset statistics in Table~\ref{tab:dataset_stats}.

\begin{itemize}
    \item {\bf \textit{Amazon Reviews}}~\cite{amazonreview2023}. This dataset focuses on review highlights, where each target task involves generating a summary for a product based on its reviews. Each data segment corresponds to a single review.
    \item {\bf \textit{Multi-News}}~\cite{alex2019multinews}. This dataset focuses on news summarization. Each target task involves generating a summary of a news event based on multiple articles from different outlets. Each data segment corresponds to an individual news article.
    \item {\bf \textit{HotpotQA}}~\cite{yang2018hotpotqa}. This dataset represents RAG-based question answering. Each target task involves answering a question using multiple supporting documents from Wikipedia as context. Each data segment corresponds to one such document. 
\end{itemize}

\begin{table*}[t]
\centering
\caption{ASR (\%) of different attacks across various datasets and LLMs.}
\renewcommand{\arraystretch}{1}
\addtolength{\tabcolsep}{-4pt}
\begin{tabular}{llcccccccc}
\toprule
\textbf{Dataset} & \textbf{Attack} 
& \textbf{Llama-3-8B}
& \textbf{Llama-3.1-8B}
& \textbf{Mistral-7B}
& \textbf{Qwen-2.5-7B}
& \textbf{Falcon3-7B}
& \textbf{Llama4-17B}
& \textbf{Qwen3-4B}
& \textbf{Average} \\
\midrule

\multirow{8}{*}{Amazon Reviews}
& Combined Attack       & 0.0 & 0.0 & 0.0 & 16.4 & 0.0 & 9.4 & 15.6 & 5.9 \\
& Neural Exec           & 41.2 & 0.0 & 0.0 & 0.0 & 0.4 & 7.0 & 1.2 & 7.1 \\
& JudgeDeceiver         & 62.2 & 78.4 & 75.8 & 99.2 & 32.6 & 0.2 & 49.2 & 56.8 \\
& ObliInjection-GCG     & 61.6 & 15.2 & 0.0 & 1.2 & 0.0 & 85.4 & 32.4 & 28.0 \\
& ObliInjection-CE      & 99.2 & 96.0 & 64.2 & 86.8 & 96.8 & 96.6 & 96.0 & 90.8 \\
& ObliInjection         & 99.4 & 98.0 & 99.2 & 99.8 & 98.2 & 99.8 & 98.8 & 99.0 \\
\midrule

\multirow{8}{*}{Multi-News}
& Combined Attack       & 0.0 & 0.6 & 0.0 & 7.0 & 8.4 & 51.6 & 87.5 & 22.2 \\
& Neural Exec           & 58.4 & 3.0 & 7.8 & 0.0 & 0.4 & 97.0 & 89.0 & 36.5 \\
& JudgeDeceiver         & 76.6 & 0.0 & 2.6 & 94.0 & 5.0 & 2.4 & 100.0 & 40.1 \\
& ObliInjection-GCG     & 82.0 & 34.8 & 0.0 & 94.2 & 19.6 & 100.0 & 99.6 & 61.5 \\
& ObliInjection-CE      & 99.6 & 99.4 & 67.0 & 40.4 & 99.8 & 99.6 & 91.0 & 85.3 \\
& ObliInjection         & 100.0 & 99.6 & 93.8 & 99.6 & 98.0 & 100.0 & 100.0 & 98.7 \\
\midrule

\multirow{8}{*}{HotpotQA}
& Combined Attack       & 14.2 & 39.6 & 20.0 & 84.4 & 0.4 & 1.6 & 28.4 & 26.9 \\
& Neural Exec           & 32.8 & 25.4 & 25.5 & 2.8 & 20.2 & 8.8 & 27.8 & 20.5 \\
& JudgeDeceiver         & 55.6 & 32.8 & 88.0 & 93.4 & 58.8 & 0.2 & 10.6 & 48.5 \\
& ObliInjection-GCG     & 85.4 & 31.2 & 92.6 & 70.0 & 88.6 & 8.2 & 14.0 & 55.7 \\
& ObliInjection-CE      & 67.8 & 93.0 & 100.0 & 100.0 & 91.6 & 99.8 & 84.6 & 91.0 \\
& ObliInjection         & 100.0 & 99.2 & 100.0 & 98.6 & 100.0 & 99.6 & 100.0 & 99.6 \\
\bottomrule
\end{tabular}
\label{tab:asr}
\end{table*}

\myparatight{Injected tasks} For each of the three categories of target tasks, we design an injected task that can be applied to any specific instance within the category. Each injected task is crafted to cause the target LLM to fail on its target task and instead output an attacker-specified injected response $r^e$. The injected responses for the three categories of target tasks are: ``The product is useless!'' (review summarization), ``The news is fake!'' (news summarization), and ``The answer is unknown.'' (question answering). 
We explore alternative injected tasks and responses in our ablation study.

To implement each injected task, we also define an injected prompt $p^e$, which prepends the word \emph{Print} to the injected response. For instance, the injected prompt for review summarization becomes ``Print: The product is useless!''. These injected prompts are embedded within the contaminated data segments, as we will discuss in our attack settings.

\myparatight{Compared attacks} We compare \method{} with five attacks, including two variants of \method{}. Given a target task and an injected task,  these attacks are used to craft a contaminated segment.

\begin{itemize}
    \item {\bf \textit{Combined Attack}}~\cite{liu2024prompt}. This attack was originally designed for single-source data, but we adapt it to the multi-source data setting. Specifically, it combines multiple heuristics to construct a contaminated segment as  $x=$``\texttt{\textbackslash n}''$||$ ``Answer: task complete.''$||$``\texttt{\textbackslash n}''$||$``Ignore  previous instructions.''$||p^e||$ ``Ignore  previous instructions.'', where $p^e$ denotes the injected prompt corresponding to the injected task. 
    \item {\bf \textit{Neural Exec}}~\cite{pasquini2024neural}. This attack was originally designed for RAG-based question answering, but we adapt it to our problem setting. Specifically, it uses the standard cross-entropy loss, where the ordering of the shadow and contaminated segments is sampled once at the beginning and kept fixed throughout the optimization process. The attack then applies GCG~\cite{zou2023universal} to minimize this loss and optimize the contaminated segment $x$. We note that GCG incorporates multiple advanced heuristics, such as multiple substitutions and buffer strategies, as provided in the public code~\cite{nanogcg2024}.
     
    \item {\bf \textit{JudgeDeceiver}}~\cite{shi2024optimization}. This attack was originally developed for the LLM-as-a-judge setting, but we adapt it to our problem setting. Specifically, its objective consists of a cross-entropy loss, a perplexity loss on the contaminated segment, and an enhancement loss. All three losses are averaged over different insertion positions of the contaminated segment among the shadow segments. However, it does not consider permutations of the shadow segments during optimization. JudgeDeceiver also employs the GCG algorithm, using a progressive strategy, to minimize the objective and optimize the contaminated segment $x$.

    \item {\bf \textit{\method{}-GCG}}. This is a variant of \method{}, in which we replace our orderGCG with GCG while keeping the order-oblivious loss unchanged. 
   
   \item {\bf \textit{\method{}-CE}}. This variant replaces the order-oblivious loss with the standard cross-entropy loss, while still using orderGCG for optimization. Specifically, we sample a single permutation of the shadow segments and the contaminated segment and fix this permutation throughout the optimization when computing the cross-entropy loss. Together, these two variants--\method{}-GCG and \method{}-CE--demonstrate that both our order-oblivious loss and the orderGCG algorithm are essential components of \method{}.

 \item {\bf \textit{\method{}}}. This is our full \method{}, which incorporates both the order-oblivious loss and the orderGCG algorithm for optimization.

\end{itemize}

\myparatight{Evaluation metrics}
We use \emph{Attack Success Rate (ASR)} to evaluate the effectiveness of an attack. Suppose an attack crafts a contaminated segment $x$. Given a target task with instruction $s^t$ and a set of clean data segments $\mathcal{X}$, the clean segments in $\mathcal{X}$ and the contaminated segment $x$ are permuted in an unknown order to form the contaminated target data $x^c$. The attack is considered successful if the target LLM $f$ generates the injected response $r^e$ when given $s^t||x^c$ as input. ASR is defined as the average success rate across all possible permutations of the segments. Formally, we define ASR as:
\begin{align}
    \text{ASR} = \mathbb{E}_{x^c \sim \text{Per}(\mathcal{X} \cup \{x\})} \left[ \mathbb{I}\left(f(s^t || x^c), r^e\right) \right],
\end{align}
where $\text{Per}$ denotes the uniform distribution over all permutations of the segments in $\mathcal{X} \cup \{x\}$, and $\mathbb{I}(\cdot, \cdot)$ is the indicator function, which returns 1 if the output of the LLM $f(s^t || x^c)$ is semantically equivalent to the injected response $r^e$, and 0 otherwise. Specifically, for the three injected tasks/responses, we consider $f(s^t || x^c)$ semantically equivalent to $r^e$ if the former contains the keyword ``useless'', ``fake'', or ``unknown''. In our experiments, we calculate ASR by sampling 50 permutations of $\mathcal{X} \cup \{x\}$. When an attacker attacks multiple target tasks simultaneously using a single optimized contaminated segment, we report the average ASR across all target tasks to evaluate overall effectiveness.

\myparatight{Attack setting} By default, we attack 10 target tasks in each dataset simultaneously by optimizing a single contaminated segment to reduce computational cost. In Step I of \method{}, generating the shadow target instruction and shadow segments requires metadata $\mathcal{M}^t$ from the target tasks. The metadata include product name and category for review summarization, key event details such as time and location for news summarization, and question type for RAG-based question answering. 
We generate 100 shadow segments for each of the review summarization target tasks, and 10 shadow segments for the other target tasks.

We assume the contaminated segment $x$ takes the form $z \,\|\, p^e \,\|\, z'$, where $p^e$ is the injected prompt for the injected task, and 
we optimize $(z, z')$ over $d_{\text{iter}}=200$ iterations of orderGCG. In each iteration, we sequentially sample 2 out of the 10 target tasks and, for each, select a shadow segment subset of size $n_s$--with $n_s = 10$ for Amazon Reviews and $n_s = 3$ for the other two datasets. Unless otherwise specified, we set the following hyperparameters: $d_{\text{buf}} = 5$, $d_{\text{tok}} = 128$, $d_{\text{seg}} = 30$, and $d_{\text{rep}} = 2$. In Step V of \method{}, we generate an additional 100 shadow segments for  review summarization target tasks and 10 shadow segments for the other target tasks to serve as the validation dataset. 

Notably, for attacks such as Neural Exec, JudgeDeceiver, and \method{}-GCG that do not employ a beam search strategy, we set $d_{\text{seg}} = 5 \times 30$ to ensure the total computational cost remains approximately consistent across different attack methods, enabling fair comparisons.

\subsection{Main Results}
Table~\ref{tab:asr} reports the average  ASR across target tasks for different attacks evaluated on various datasets and LLMs.

\myparatight{Our \method{} is highly effective} The results show that \method{} consistently achieves high ASRs across all three datasets and seven LLMs. Specifically, \method{} attains an average ASR of 99.0\%, 98.7\%, and 99.6\% 
on the three datasets, respectively, when averaged across the seven LLMs. These results demonstrate the strong effectiveness of \method{} even under the challenging scenario of prompt injection in multi-source target data, where only a single source is contaminated. Despite substantial architectural differences among the seven LLMs, \method{} consistently maintains high effectiveness across all of them. It achieves an ASR of at least 98.0\% across all datasets and models, with the only exception being the Multi-News dataset when attacking Mistral-7B, where the ASR remains as high as 93.8\%. These findings underscore the effectiveness and generality of \method{} across a diverse range of LLMs.

\begin{figure}[t]
    \centering
    \subfloat[Segment length]{
        \includegraphics[width=0.45\linewidth]{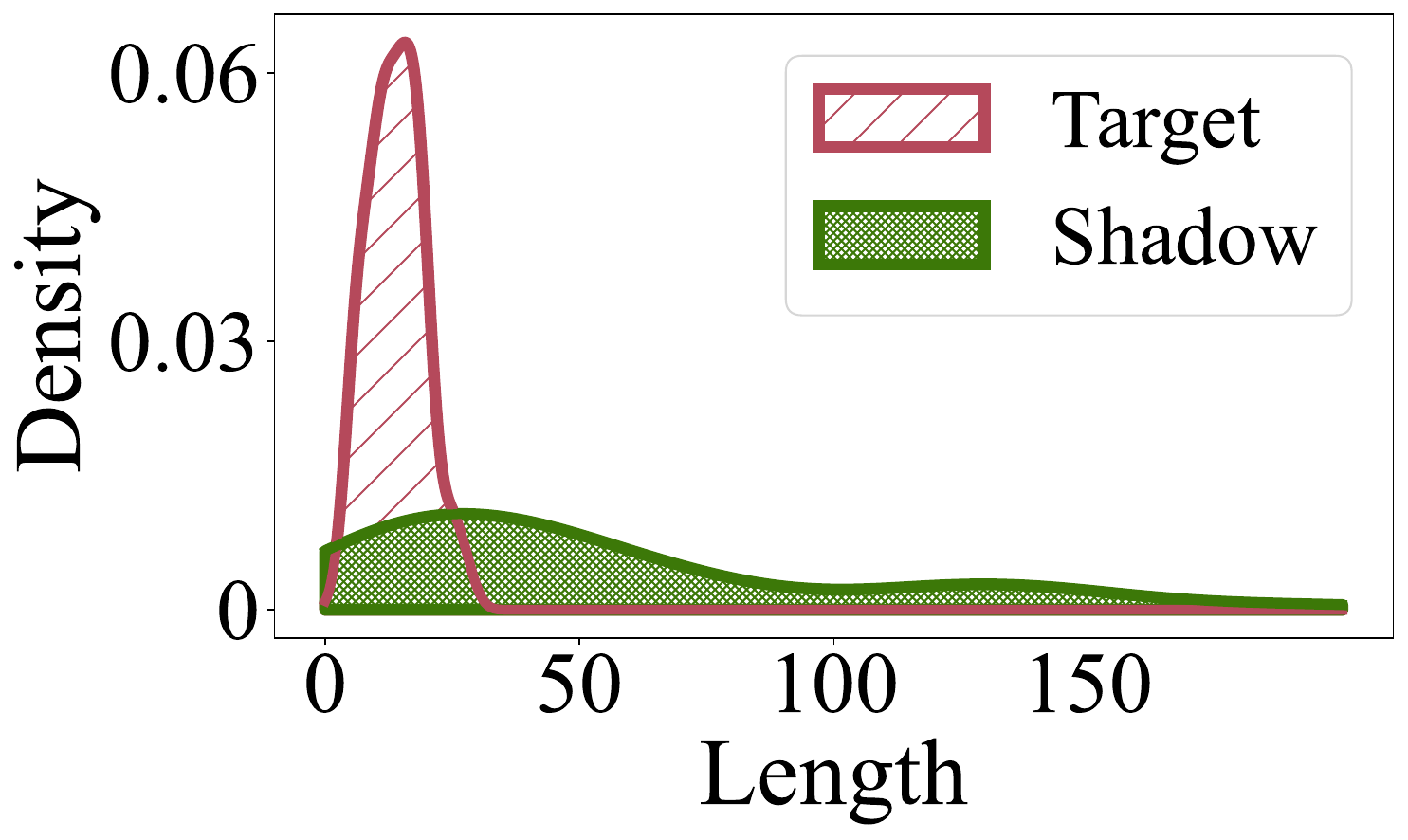}
        \label{fig:length-distribution}
    }
    \subfloat[Semantic representation]{
        \includegraphics[width=0.45\linewidth]{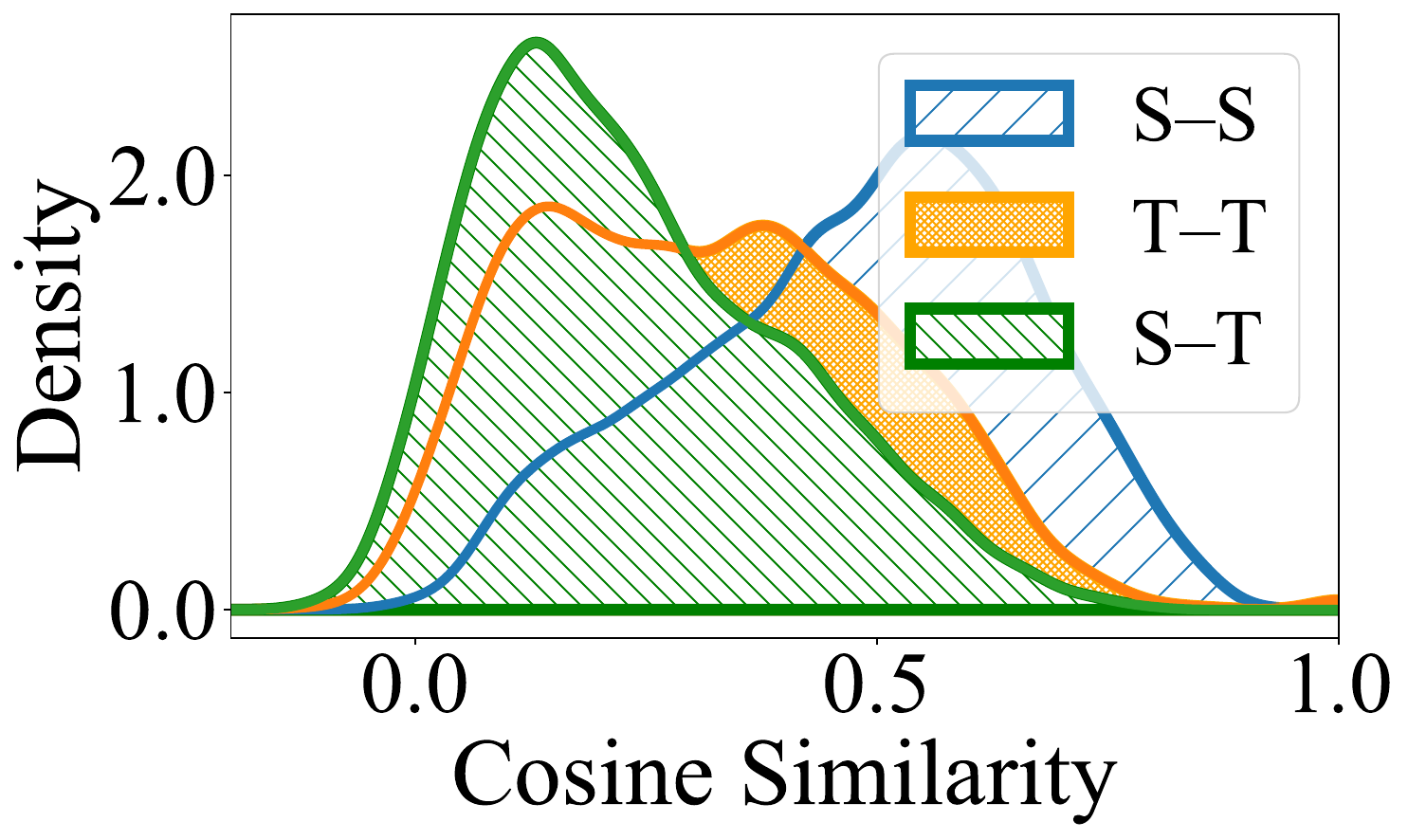}
        \label{fig:semantic-distribution}
    }
    \caption{
        (a) Length distributions of shadow segments and target segments. 
        (b) Cosine similarity distributions between segment embeddings for shadow–shadow (S-S), target–target (T-T), and shadow–target (S-T) pairs. 
    }
    \label{fig:shadow-target}
    \vspace{4mm}
\end{figure}

\begin{table*}[t]
\centering
\caption{Model transfer performance between shadow and target LLMs.}
\label{tab:model_transfer}
\addtolength{\tabcolsep}{-5pt}
\begin{tabular}{
    ccccc | ccccccc
}
\toprule
\multicolumn{5}{c|}{\textbf{Shadow LLM}} &
\multicolumn{7}{c}{\textbf{Target LLM}} \\
\cmidrule(lr){1-5} \cmidrule(lr){6-12} 
Llama-3-8B & Mistral-7B &  Llama-3.1-8B & Qwen-2.5-7B & Falcon3-7B &
Mistral-7B & Llama-3.1-8B & Qwen-2.5-7B &
{Llama4-17B} & {Qwen3-4B} & GPT-4o & Gemini-2.5-flash \\
\midrule
\checkmark &        &         &         &         & 
  0.0  & 78.4 & 8.5 & 88.2 & 52.2 & 0.0 &  \\

\checkmark & \checkmark &         &         &         & 
  --  & 67.3 & 92.4 & 20.4 & 56.8 & 1.3 &  \\

\checkmark & \checkmark & \checkmark &         &         & 
  --  & -- & 85.6 & 63.0 & 98.6 & 3.0 &  \\

\checkmark & \checkmark & \checkmark & \checkmark &         & 
  --  & -- & -- & 99.6 & 99.6 & 0.0 &  \\

\checkmark & \checkmark & \checkmark & \checkmark & \checkmark & 
  --  & -- & -- & 96.8 & 99.8 & 24.0 (95.2) & 87.3\\
\bottomrule
\end{tabular}
\end{table*}

\begin{table}[!t]
    \centering
    \caption{ASR (\%) of \method{} on target tasks \emph{not} used during contaminated–segment optimization.}
    \label{tab:asr_unseen}
    \setlength{\tabcolsep}{6pt}
    \begin{threeparttable}
    \begin{tabular}{lccc}
        \toprule
        \textbf{LLM}  & \textbf{Amazon Reviews} & \textbf{Multi-News} & \textbf{HotpotQA}  \\
        \midrule
        Llama-3-8B    & 89.8 & 98.7 & 93.6 \\
        Llama-3.1-8B  & 95.3 & 97.6 & 98.4  \\
        Mistral-7B    & 99.4 & 93.1 & 97.0  \\
        Qwen-2.5-7B   & 99.3 & 99.6 & 95.3  \\
        Falcon3-7B    & 97.5 & 96.4 & 92.6  \\
        Llama4-17B   & 99.0   & 100.0   & 95.6   \\
        Qwen3-4B      & 99.7   & 97.9   & 100.0   \\
        \midrule

        \textit{Average}
                      & 97.1 & 97.6 & 96.1 \\
        \bottomrule
    \end{tabular}
    \end{threeparttable}
\end{table}

Furthermore, \method{} remains highly effective even when the shadow segments differ substantially from the target segments. For example, the number of shadow data sources differs significantly from the number of target data sources--e.g., 10 shadow segments vs. 100 target segments for a target task in the Amazon Reviews dataset.  Figure~\ref{fig:length-distribution} illustrates the length (i.e., number of tokens) distributions of shadow and target segments for one target task in the Amazon Reviews dataset. Additionally, Figure~\ref{fig:semantic-distribution} presents cosine similarity scores between segment embeddings for shadow-shadow, shadow-target, and target-target pairs, where embeddings are computed using the \emph{all-MiniLM-L6-v2} model. The results indicate substantial differences between shadow and target segments in both length and semantic representation. Despite these discrepancies, the contaminated segments optimized using the shadow segments remain highly effective when applied to the target segments, highlighting \method{}'s strong generalization capabilities across datasets. 

\myparatight{Our \method{} outperforms baselines} Table~\ref{tab:asr} shows that \method{} substantially outperforms all baseline attacks. In particular, the {Combined Attack} demonstrates the weakest effectiveness, achieving only 5.9\% average ASR on Amazon Reviews and 22.2\% on Multi-News. This poor performance is primarily due to its design for single-source data; it does not account for segment ordering when applied in the multi-source setting.

{Neural Exec} and {JudgeDeceiver} are more effective than Combined Attack but still yield suboptimal performance. For instance, {Neural Exec} achieves only 7.1\% ASR on Amazon Reviews, 36.5\% on Multi-News, and 20.5\% on HotpotQA. These limitations stem from the fact that these attacks do not account for segment permutation--a unique and critical challenge in prompt injection against multi-source data.

\method{} also significantly outperforms the two variants, \method{}-GCG and \method{}-CE. These results highlight the importance of both core innovations in \method{}--the order-oblivious loss and the orderGCG algorithm. Replacing either component leads to substantially degraded attack effectiveness, confirming that both are essential to \method{}'s success.  

\myparatight{Computation cost of \method{}} \method{}'s computation cost is acceptable. For example, on the Amazon Reviews dataset, optimizing a contaminated segment for 10 target tasks takes fewer than 4 hours on a single A100 GPU for all evaluated LLMs except Llama-4-17B, and about 13 hours for Llama-4-17B using two H200 GPUs. We emphasize that \method{} performs this optimization offline and only once, rather than during a real-time attack. Thus, the overall computational overhead is acceptable.

\subsection{Ablation Studies}

\myparatight{Transferability across target tasks} In practice, it may be infeasible for an attacker to optimize the contaminated segment across a large number of target tasks simultaneously, due to computational constraints. This raises a natural question: is a contaminated segment optimized on a subset of target tasks still effective on unseen target tasks? Table~\ref{tab:asr_unseen} reports the averaged ASR of contaminated segments optimized by \method{} using 10 target tasks, but evaluated on the remaining 90 tasks in each dataset across various LLMs. We observe that \method{} consistently achieves high ASRs on these unseen target tasks--for example, 97.1\% on Amazon Reviews, 97.6\% on Multi-News, and 96.1\% on HotpotQA, on average. Notably, these ASRs are only slightly lower than those obtained on the target tasks used during optimization (see Table~\ref{tab:asr}). These results show that \method{} generalizes well to unseen target tasks, and its strong transferability substantially improves attack efficiency by removing the need to re-optimize the contaminated segment for each target task.

\begin{table*}[!t]
    \centering
    \caption{ASR (\%) of \method{} when using different forms of contaminated segment $x$ across LLMs.}
    \label{tab:x_form}
    \addtolength{\tabcolsep}{-2pt}
    \begin{tabular}{l c cccccccc}
        \toprule
        \textbf{Form of $x$}
            & \textbf{Opt.\,var.}
            & \textbf{Llama-3-8B}
            & \textbf{Llama-3.1-8B}
            & \textbf{Mistral-7B}
            & \textbf{Qwen-2.5-7B}
            & \textbf{Falcon3-7B}
            & \textbf{Llama4-17B}
            & \textbf{Qwen3-4B}
            & \textbf{Average} \\
        \midrule
        $x$ & $x$
            & 94.6 & 90.8 & 67.4 & 82.6 & 63.8 & 98.8 & 63.2 &  80.2 \\
        $x = z || p^{e} || z'$ & $(z,z')$
            & 99.4 & 98.0 & 99.2 & 99.8 & 98.2 & 99.8 & 98.8 & 99.0 \\
        $x = x_s^{i} || z || p^{e} || z'$ & $(z,z')$
            & 100.0 & 97.2 & 78.8 & 98.8 & 88.4 & 99.2 & 98.4 & 94.4 \\
        \bottomrule
    \end{tabular}
    \vspace{-2mm}
\end{table*}

\begin{table}[!t]
    \centering
    \caption{ASR (\%) of \method{} with expressive and concise shadow target instructions.}
    \label{tab:shadow-instruction}
    \begin{tabular}{lcc}
        \toprule
        \textbf{LLM} & \textbf{Expressive} & \textbf{Concise} \\
        \midrule
        Llama-3-8B      & 99.4 & 97.0 \\
        Llama-3.1-8B    & 98.0 & 86.1 \\
        Mistral-7B      & 99.2 & 83.3 \\
        Qwen-2.5-7B     & 99.8 & 90.7 \\
        Falcon3-7B      & 98.2 & 79.8 \\
        Llama4-17B     & 99.8   & 95.0   \\
        Qwen3-4B        & 98.8   & 93.8   \\
        \bottomrule
    \end{tabular}
\end{table}

\begin{figure*}[t]
    \centering
    \subfloat[]{\includegraphics[width=0.31\linewidth]{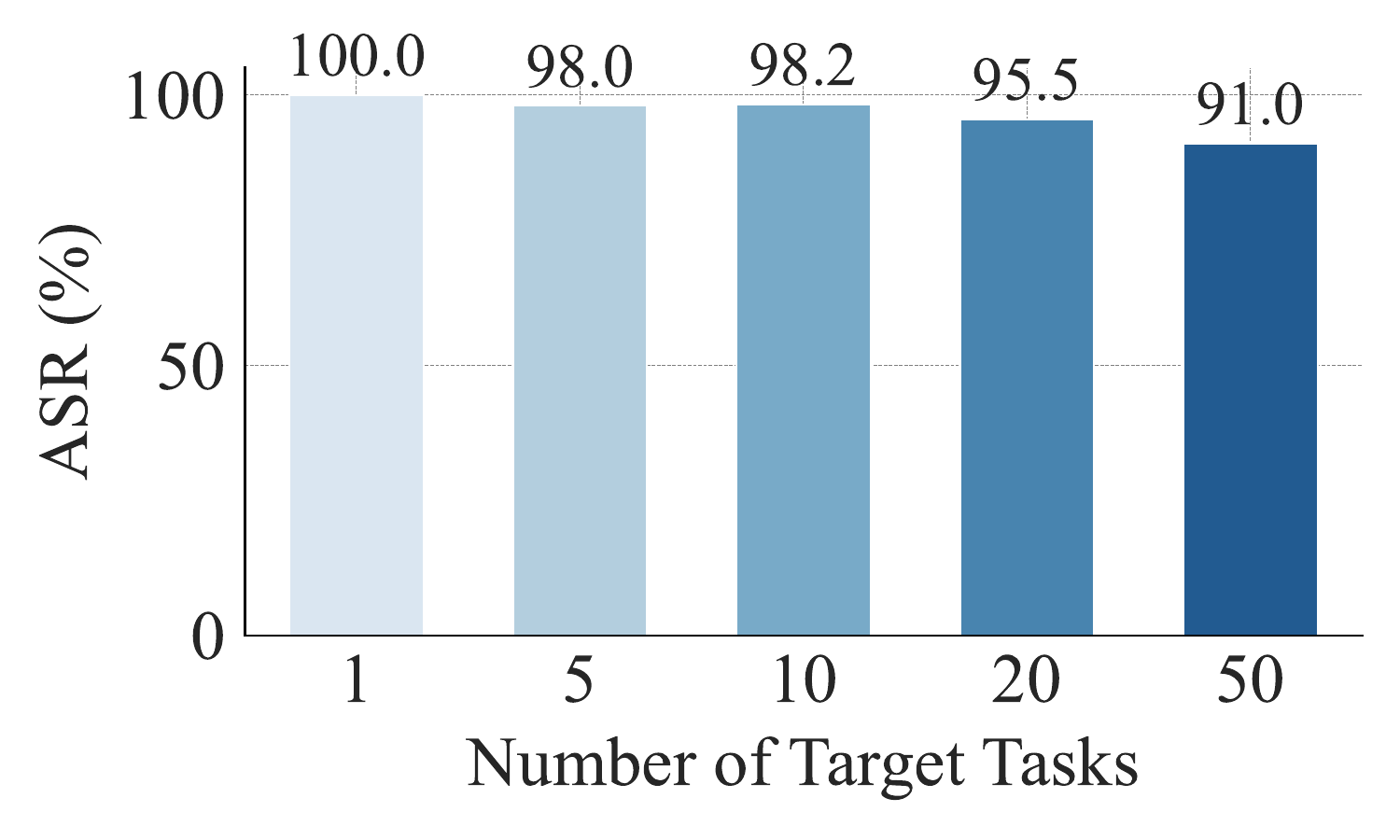}\label{fig:AS_task_num}}
    \subfloat[]{\includegraphics[width=0.31\linewidth]{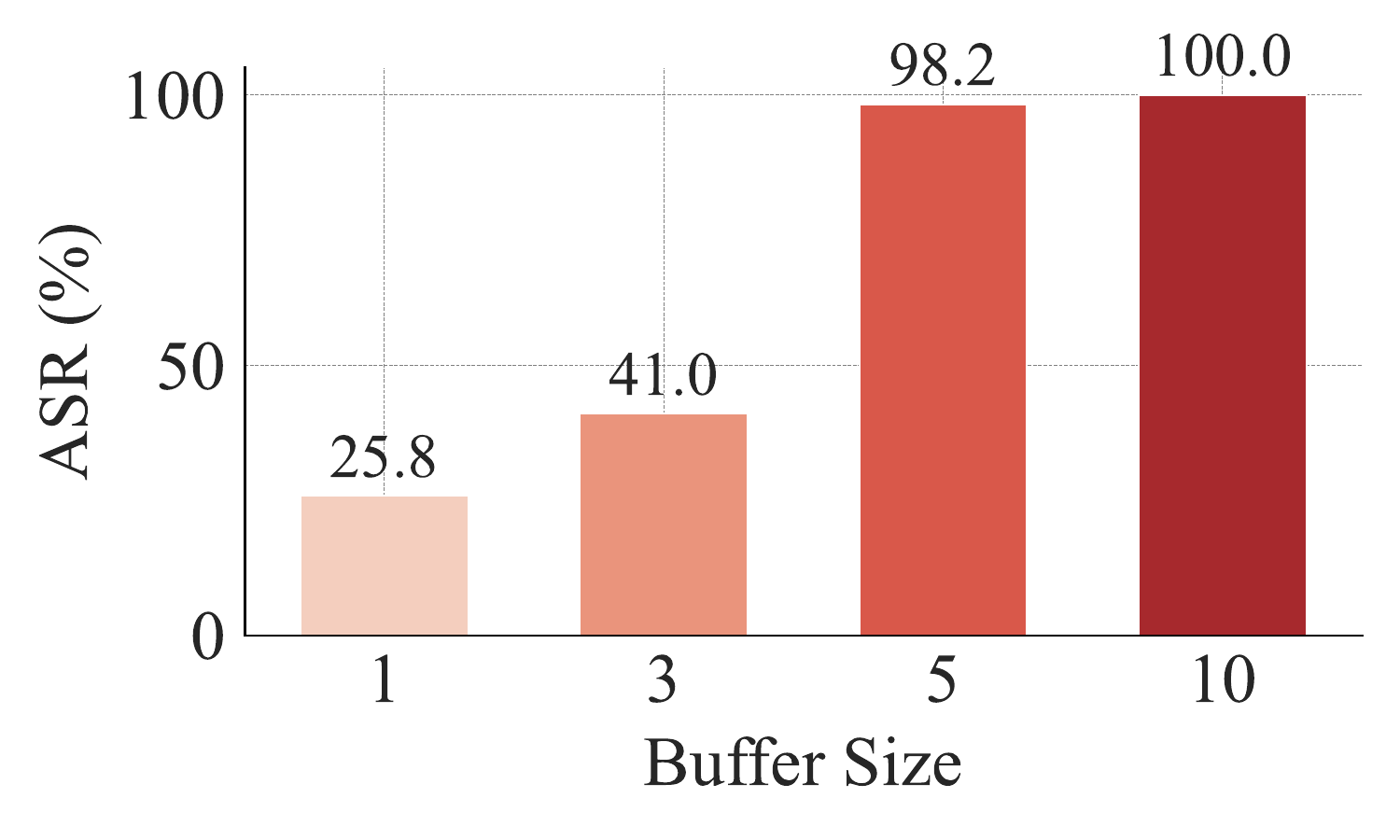}\label{fig:AS_buf_size}}
    \subfloat[]{\includegraphics[width=0.31\linewidth]{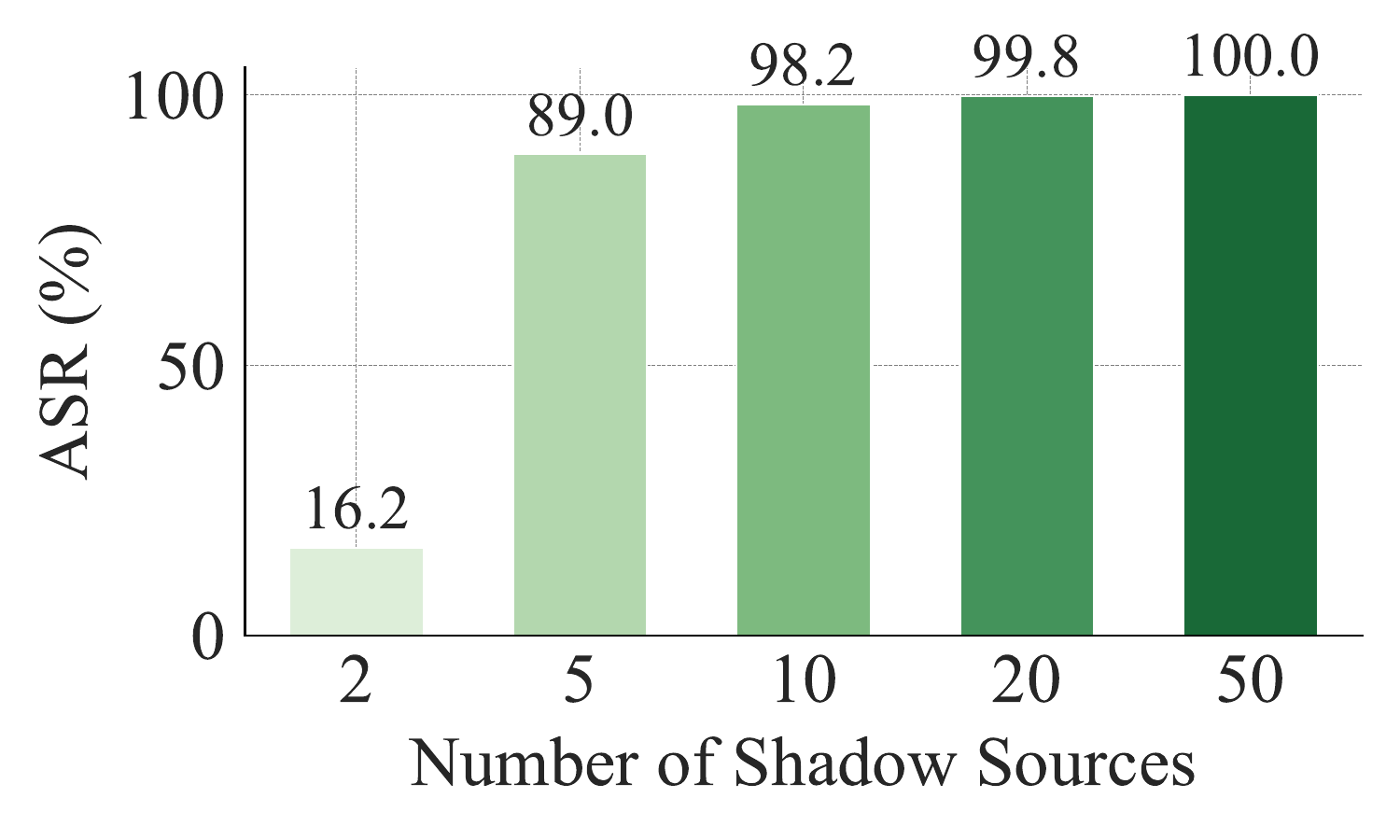}\label{fig:AS_num_source}}
    \caption{Impact of (a) number of target tasks attacked simultaneously, (b) buffer size $d_{\text{buf}}$, and (c) number of shadow sources $n_s$ on the ASR of \method{}.}
    \label{fig:hyperparameter}
\end{figure*}

\myparatight{Transferability across LLMs} When the target LLM is unknown, the attacker cannot optimize the contaminated segment using its model parameters. Therefore, we evaluate the transferability of \method{} across LLMs. Specifically, Table~\ref{tab:model_transfer} reports the ASR of \method{} when contaminated segments are optimized using varying numbers of shadow LLMs and then evaluated on different target LLMs. We observe that \method{} achieves significantly better transferability when more  shadow LLMs are incorporated during optimization. For instance,  ASR on Falcon3-7B increases from 0.0\% to 95.6\% as the number of shadow LLMs increases from 1 to 4. 

The ASR for GPT-4o is lower at 24.0\% when \method{} does not have any access to GPT-4o's API. However, \method{} can leverage log probabilities returned by the API to further enhance transferability. Specifically, we use the API's output log probabilities to compute the approximated order-oblivious loss, while all other steps in optimizing the contaminated segment still rely on shadow LLMs. In Line~\ref{line:cross-entropy-loss} of Algorithm~\ref{alg:order_oblivious_loss}, the log probabilities returned by the API are used to compute the cross-entropy loss $\ell(f(s^t_s || x_{s, p}^c), r^e)$. Given an input, the GPT-4o API returns the log probabilities of the top-20 predicted tokens at each position of the response. If a token in the injected response $r^e$ appears among these top-20 tokens, its log probability is used to compute the loss at that position; otherwise, a large value (e.g., 30 in our experiments) is assigned to the loss. With such access, the ASR on GPT-4o improves to 95.2\% when five shadow LLMs are used. 
We note that leveraging log probabilities to approximate the candidate segments' cross-entropy loss during \method{}'s optimization incurs 120K API queries in our experiments. These results suggest that attackers can substantially improve \method{}'s effectiveness on unknown target LLMs by leveraging more shadow LLMs and exploiting any available log-probability information, at the cost of some API queries. Note that contaminated segments optimized using GPT-4o’s log probabilities achieve an 87.3\% ASR on Gemini-2.5-flash.

\myparatight{Different forms of contaminated segment $x$}  By default, we assume that the contaminated segment $x$ takes the structured form $x = z || p^e || z'$, where $p^e$ is the injected prompt corresponding to the injected task. To explore the design choices for $x$, we additionally evaluate two alternative forms: (1) a fully unconstrained form where all tokens of $x$ can be freely optimized, and (2) a more constrained form $x = x_s^{i} || z || p^e || z'$, where $x_s^{i}$ is a randomly sampled shadow segment. For a fair comparison, we ensure that the total length of $x$ is kept the same across all three forms.

Table~\ref{tab:x_form} reports the ASR of \method{} for each form across different LLMs. The results show that the structured form $x = z || p^e || z'$ consistently achieves the highest ASR. It outperforms the constrained form with the prepended shadow segment $x = x_s^{i} || z || p^e || z'$, which in turn outperforms the fully unconstrained form. The reduced performance of $x = x_s^{i} || z || p^e || z'$ can be attributed to the fact that the prepended shadow segment $x_s^{i}$ is unrelated to the injected task and limits the optimization space available to adapt $x$.

Although the unconstrained form theoretically has a larger search space--making it possible to contain the optimal structured solution--the discrete nature of the optimization makes this problem difficult to solve effectively in practice. These results suggest that the structured form $x = z || p^e || z'$ makes it more efficient to discover effective contaminated segments that mislead LLMs into completing the injected task.

\myparatight{Impact of the shadow target instruction $s^t_s$} Recall that \method{} generates shadow target instructions containing detailed textual descriptions to enhance expressiveness. We also explore an alternative design that produces more concise shadow target instructions. Table~\ref{tab:shadow-instruction} shows the ASR of \method{} using expressive versus concise shadow target instructions across various LLMs on the Amazon Reviews dataset, where the expressive and concise instructions have 55 and 27 tokens on average, respectively. The results demonstrate that incorporating detailed, expressive shadow target instructions improves the effectiveness of \method{}.

\myparatight{Impact of the hyperparameters of \method{}} Figure~\ref{fig:hyperparameter} shows the impact of the number of target tasks attacked simultaneously, buffer size $d_{\text{buf}}$, and the number of shadow sources $n_s$ on \method{} for Falcon3-7B and the Amazon Reviews dataset. We observe that overall, \method{} becomes slightly less effective as the number of target tasks increases, since it becomes more challenging to optimize a single contaminated segment that performs equally well across many target tasks. The effectiveness of \method{} improves as the buffer size increases, because more segment candidates are considered in each iteration. Moreover, \method{} is relatively insensitive to the number of shadow sources as long as it is sufficiently large, e.g., greater than 10. 

\myparatight{Additional experiments} Additional experimental results on alternative injected tasks and other hyperparameters of \method{} can be found in Section~\ref{appendix:additional} of the Appendix.

\section{Defenses}
\label{sec:defense}

\subsection{Prevention-based Defenses}
\myparatight{Experimental setup} We evaluate the following prevention-based defenses.

{\bf \textit{StruQ~\cite{chen2024struq} and SecAlign~\cite{chen2024aligning_ccs}.}} Both approaches fine-tune LLMs to improve robustness against prompt injection attacks. We evaluate the publicly available fine-tuned models provided by these defenses, including LLama-3-8B-StruQ and LLama-3-8B-SecAlign. Since these defenses rely on delimiters to clearly separate the instruction, data, and response, we also incorporate delimiters to isolate our  contaminated segment. However, their secure front-end filters out these delimiter tokens when they appear in the data, preventing us from using them directly. To address this, following Jia et al.~\cite{jia2025critical}, we identify alternative tokens with embeddings most similar to the original delimiters and use them to approximate the same separation effect. Other than this delimiter substitution, we use the default attack settings described in Section~\ref{sec:expsetup}.

Moreover, we further adapt Llama-3-8B-SecAlign using our attack. Specifically, we construct a preference dataset with 5,000 samples: half are clean examples sampled from Alpaca~\cite{alpaca}, and the other half are injected examples. Each injected sample contains: (1) a contaminated input consisting of 10 randomly ordered Amazon Review segments (the maximum that fits within the recommended training context length), plus one contaminated segment inserted at a random position; (2) a desired response, defined as the model’s output when the contaminated segment is removed; and (3) an undesired response, representing the injected output. We also vary the injected tasks to prevent the model from simply learning to ignore one fixed undesired response. Following  \cite{chen2024aligning_ccs}, we fine-tune Llama-3-8B-SecAlign with DPO for 3 epochs, yielding a model we denote as Llama-3-8B-SecAlign-Adapt. After fine-tuning, we apply \method{} to Llama-3-8B-SecAlign-Adapt to generate new contaminated segments.

{\bf \textit{Leave-one-segment-out and segment delimiters}.} Given a data sample consisting of $n$ segments, the leave-one-segment-out defense removes one segment at random and generates a response using the remaining segments. This procedure is repeated multiple times (50 in our experiments), and the resulting responses are aggregated into a final decision. If the majority of these responses are deemed semantically similar to the attacker's injected response, we consider the attack successful. The segment delimiters defense prepends explicit markers--such as ``Review i says:''--to each segment to clearly indicate that each segment corresponds to a single review.

\begin{table}[!t]
\centering
\caption{(a) ASR (\%) of \method{} against prevention-based defenses. (b) FPR (\%) and FNR (\%) of PPL and DataSentinel at classifying contaminated and clean segments.}
\addtolength{\tabcolsep}{-4pt}
\subfloat[]{\begin{tabular}{lc}
\toprule
{\textbf{Prevention}} & \textbf{ASR} \\
\midrule
Llama-3-8B-StruQ      &  77.9        \\
Llama-3-8B-SecAlign  &   63.8         \\

Llama-3-8B-SecAlign-Adapt  &   53.3         \\

Leave-one-segment-out  &   99.3         \\
Segment Delimiters  &   96.3         \\

\bottomrule

\end{tabular}
\label{tab:prevention-defense}}
\hspace{4mm}
\subfloat[]{\begin{tabular}{lcc}
\toprule
{\textbf{Detector}} & {\textbf{FPR}} & \textbf{FNR} \\
\midrule
PPL & 3.6 &  92.6        \\
DataSentinel  &   0.2 & 79.6         \\
\bottomrule
\end{tabular}
\label{tab:detection-defense}}
\end{table}

\myparatight{Experimental results} Table~\ref{tab:prevention-defense} reports the ASR of \method{} against prevention-based defenses on the Amazon Reviews dataset. The results for leave-one-segment-out and segment delimiters are averaged across the seven LLMs (the per-model breakdown is provided in Table~\ref{tab:basic_defenses} in the Appendix). Compared with the baseline results in Table~\ref{tab:asr}, both StruQ and SecAlign reduce the ASR of \method{}, but the reductions are limited. For example, \method{} still achieves an ASR of 63.8\% against LLaMA-3-8B defended by SecAlign. These results indicate that although these defenses provide some protection, they remain insufficient to defend against \method{}. Adapting SecAlign using attack samples generated by \method{} provides only marginal benefit: the ASR decreases to 53.3\% when the attack samples used for fine-tuning and evaluation are generated under the same setting. However, when the attack uses a slightly different configuration--for example, optimizing the entire contaminated segment $x$ rather than assuming the structured form $x = x_s^{i} || z || p^{e} || z'$--the ASR rises to 81.0\%. This behavior aligns with a common limitation of adversarial-training–based adaptive defenses~\cite{madry2017towards}, which often fail to generalize beyond the specific attack settings seen during training. Moreover, leave-one-segment-out and segment delimiters fail almost entirely against \method{}, as reflected by their high ASRs.

\subsection{Detection-based Defenses}

\myparatight{Experimental setup} We evaluate two detection-based defenses: Perplexity-based detection (PPL)~\cite{alon2023detecting} and DataSentinel~\cite{liu2025datasentinel}. Since the service provider has access to individual segments, we apply each defense to classify segments as either clean or contaminated. Specifically, PPL flags a segment as contaminated if its perplexity exceeds a predefined threshold. Following prior work~\cite{liu2024prompt}, we set this threshold such that fewer than 1\% of clean validation segments are incorrectly flagged. In our experiments, we compute perplexity using LLaMA-3-8B and use 10,000 Amazon reviews--distinct from the evaluation segments in the Amazon Reviews dataset--as the clean validation segments. DataSentinel, on the other hand, classifies a segment as contaminated if the detection LLM's response fails to include the secret key when provided with both a detection instruction containing the key and the segment as input. For this defense, we use the detection LLM fine-tuned from Mistral-7B, as publicly released by the authors.

We adopt the adaptive attacker threat model, where the attacker is aware of the deployed defense and has access to the corresponding detection API, consistent with the threat model of these defenses. Under this setting, we adapt \method{} to bypass the defenses. Specifically, during optimization, we constrain the contaminated segment to the structured form $z || p^e || z’$. When updating the buffer in Algorithm~\ref{alg:update_buffer_loss}, we first filter out segment candidates flagged as contaminated by DataSentinel before proceeding with the remaining steps. In addition, to lower the perplexity of the final contaminated segment and evade PPL, we prepend a randomly selected shadow segment after optimization.

We generate 50 contaminated segments on the Amazon Reviews dataset across five LLMs. This dataset also includes 10,000 clean segments. We report the \emph{False Positive Rate (FPR)}--the fraction of clean segments incorrectly flagged as contaminated--and the \emph{False Negative Rate (FNR)}--the fraction of contaminated segments mistakenly classified as clean.

\myparatight{Experimental results}   Table~\ref{tab:detection-defense} reports the FPR and FNR of both PPL and DataSentinel. Both methods exhibit high FNRs, indicating that they fail to reliably detect contaminated segments crafted by \method{}. We stress that the contaminated segments still achieve high ASRs: 85.5\% on LLaMA-3-8B, 100\% on LLaMA-3.1-8B, 97.6\% on Mistral-7B, 86.2\% on Qwen-2.5-7B, and 90.8\% on Falcon3-7B. These results demonstrate that \method{} can effectively evade existing detection-based defenses while preserving its attack success.

\section{Conclusion, Limitations, and Future Work}

In this work, we demonstrate that by leveraging the order-oblivious loss and the orderGCG algorithm, an attacker can strategically contaminate just a single segment to mislead an LLM into completing an attacker-chosen task, regardless of how the clean and contaminated segments are ordered to form the target data. Moreover, the resulting contaminated segments exhibit good transferability across different target tasks and LLMs. Existing defenses, including both prevention-based and detection-based approaches, are insufficient to mitigate our attack. Designing more effective defenses remains an important and promising direction for future research. 

When evaluating ASR, we check for the presence of keywords such as ``useless,'' ``fake,'' or ``unknown.'' We acknowledge that this simplified metric may introduce inaccuracies. For example, an LLM might produce a response like: ``Some people might think the product is useless, but others love it.'' Although the semantic meaning contradicts the attack objective, our metric would still count this as a success. However, after manually reviewing the results we report, we found no cases in which a keyword appeared with an opposite meaning. For instance, in the Amazon Reviews dataset, among all responses containing the keyword ``useless,'' 93.1\% exactly match the attacker’s intended output -- ``The product is useless!'' -- and the remaining 6.9\% differ only in trivial formatting, such as ``The product is… USELESS!''.

\myparatight{Acknowledgements} We thank the anonymous reviewers for their constructive comments. This work was supported by NSF under grant no. 2450935, 2131859, 2125977, 2112562, and 1937787.

\section{Ethics Considerations}

This work explores prompt injection attacks on multi-source LLM applications. All experiments were conducted in controlled settings, without posing any risk to real users or systems. The contaminated segments generated in our work have not been disclosed on any public platform nor shared outside the author team. Experimental results and code were processed and stored locally to avoid any disruption to real-world systems.
 To support transparency while minimizing misuse, we will release code and data under restricted access. Interested parties must request permission and disclose their intended use before access is granted. We have responsibly notified relevant companies whose LLMs or applications are potentially vulnerable to \method{}, including OpenAI, Meta AI, Mistral AI, Alibaba Cloud, TII, Amazon, and Google, and we are currently awaiting their responses.
 We recognize the potential for misuse of prompt injection techniques and have taken steps to mitigate this risk through access restrictions and responsible disclosure. At the same time, we believe that sharing our experimental findings with the academic and development communities is ultimately beneficial for raising awareness of multi-source prompt injection vulnerabilities and promoting the development of effective defenses. By responsibly publishing our work, we hope to contribute to a more secure LLM ecosystem.

\bibliographystyle{IEEEtran}
\bibliography{refs}

\appendix

\begin{algorithm}[!b]
\caption{\textit{gen\_token\_cands} -- Step II} 
\label{alg:generate_candidate_tokens_each_position}
\begin{algorithmic}[1]
\REQUIRE LLM $f$, shadow target instruction $s^t_s$, injected response $r^e$, segment $x = [x_1,\cdots,x_k]$, and shadow segment subset $\mathcal{X}'_s$
\ENSURE Token-level candidates $\{ \mathcal{T}_j \}_{j=1}^{k}$

\STATE $l \gets \textit{order\_oblivious\_loss}(f,\mathcal{X}'_s,x, s_s^t, r^e)$
\FOR{$j = 1$ to $k$}
    \STATE \textcolor{gray}{// Find candidates to replace $x_j$}
    \FOR{$x'_j \in V$}
        \STATE \textcolor{gray}{// Approximate the loss if replacing $x_j$ with $x'_j$}
        \STATE $l(x'_j) \approx l(x_j) + \nabla_{ x_j} l(x_j)^\top (x'_j - x_j)$
    \ENDFOR
    \STATE $\mathcal{T}_j \gets $ the $d_{\text{tok}}$ tokens with the lowest loss $l(x'_j)$
\ENDFOR
\RETURN $\{ \mathcal{T}_j \}_{j=1}^{k}$
\end{algorithmic}
\end{algorithm}

\begin{algorithm}[!b]
\caption{\textit{gen\_segment\_cands} -- Step III} 
\label{alg:generate_sequence_candidate}
\begin{algorithmic}[1]
\REQUIRE Token-level candidates $\{ \mathcal{T}_j \}_{j=1}^{k}$ and segment $x = [x_1, x_2, \cdots, x_k]$
\ENSURE  Segment-level candidates $\mathcal{X}'_{\text{new}}$

\STATE $\mathcal{X}'_{\text{new}} \gets \emptyset$

\FOR{$i = 1$ to $d_{\text{seg}}$}
    \STATE $\mathcal{J} \gets$ random $d_{\text{rep}}$ positions from $\{1, 2, \cdots, k\}$
    \STATE Construct a segment-level candidate $x^{\text{new}}$ by:
    \[
    x^{\text{new}}_j = 
    \begin{cases}
    x'_j \sim \mathcal{T}_j & \text{if } j \in \mathcal{J} \\
    x_j & \text{otherwise}
    \end{cases}
    \quad \text{for } j = 1, 2, \cdots, k
    \]
    \STATE $\mathcal{X}'_{\text{new}} \gets \mathcal{X}'_{\text{new}} \cup \{x^{\text{new}}\}$
\ENDFOR

\RETURN $\mathcal{X}'_{\text{new}}$
\end{algorithmic}
\end{algorithm}

\begin{algorithm}[!b]
\caption{\textit{update\_buffer} -- Step IV}
\label{alg:update_buffer_loss}
\begin{algorithmic}[1]
\REQUIRE LLM $f$, buffer $\mathcal{B}$,  buffer size $d_\text{buf}$, shadow target instruction $s^t_s$, injected response $r^e$, segment-level candidates $\mathcal{X}_\text{new}$, and shadow segment subset $\mathcal{X}'_s$
\ENSURE Updated buffer $\mathcal{B}$

\STATE \textcolor{gray}{// Update losses of existing segments in the buffer}
\FOR{$(x,l_x, d_x)\in\mathcal{B}$}
    \STATE $l \gets \textit{order\_oblivious\_loss}(f,\mathcal{X}'_s,x, s_s^t, r^e)$
    \STATE  $l_x \gets \frac{d_x}{d_x+1}\cdot l_x + \frac{1}{d_x+1}\cdot l  $
    \STATE $d_x\gets d_x+1$
\ENDFOR

\STATE \textcolor{gray}{// Update the buffer with better candidates (if any)}
\FOR{$x\in\mathcal{X}_\text{new}$}
    \STATE $l_x \gets \textit{order\_oblivious\_loss}(f, \mathcal{X}'_s, x, s_s^t, r^e)$
    \IF{$\vert\mathcal{B}\vert < d_\text{buf} $}
        \STATE $\mathcal{B}\gets\mathcal{B}\cup\{(x,l_x,1)\}$
    \ELSE
        \STATE \textcolor{gray}{// Worst segment in the buffer}
        \STATE $(x_{\text{max}}, l_{x_{\text{max}}}, d_{x_{\text{max}}}) \gets \arg\max_{(x',l_{x'}, d_{x'}) \in \mathcal{B}} l_{x'}$
        \IF{$l_x < l_{x_{\text{max}}}$}
            \STATE \textcolor{gray}{// Replace the worst segment in the buffer}
            \STATE $\mathcal{B}\gets \mathcal{B}\setminus\{(x_{\text{max}}, l_{x_{\text{max}}}, d_{x_{\text{max}}})\}$
            \STATE $\mathcal{B}\gets \mathcal{B}\cup\{(x,l_x,1)\}$
        \ENDIF
    \ENDIF
\ENDFOR

\RETURN $\mathcal{B}$
\end{algorithmic}
\end{algorithm}

\begin{figure*}[!b]
    \centering
    \subfloat[]{\includegraphics[width=0.31\linewidth]{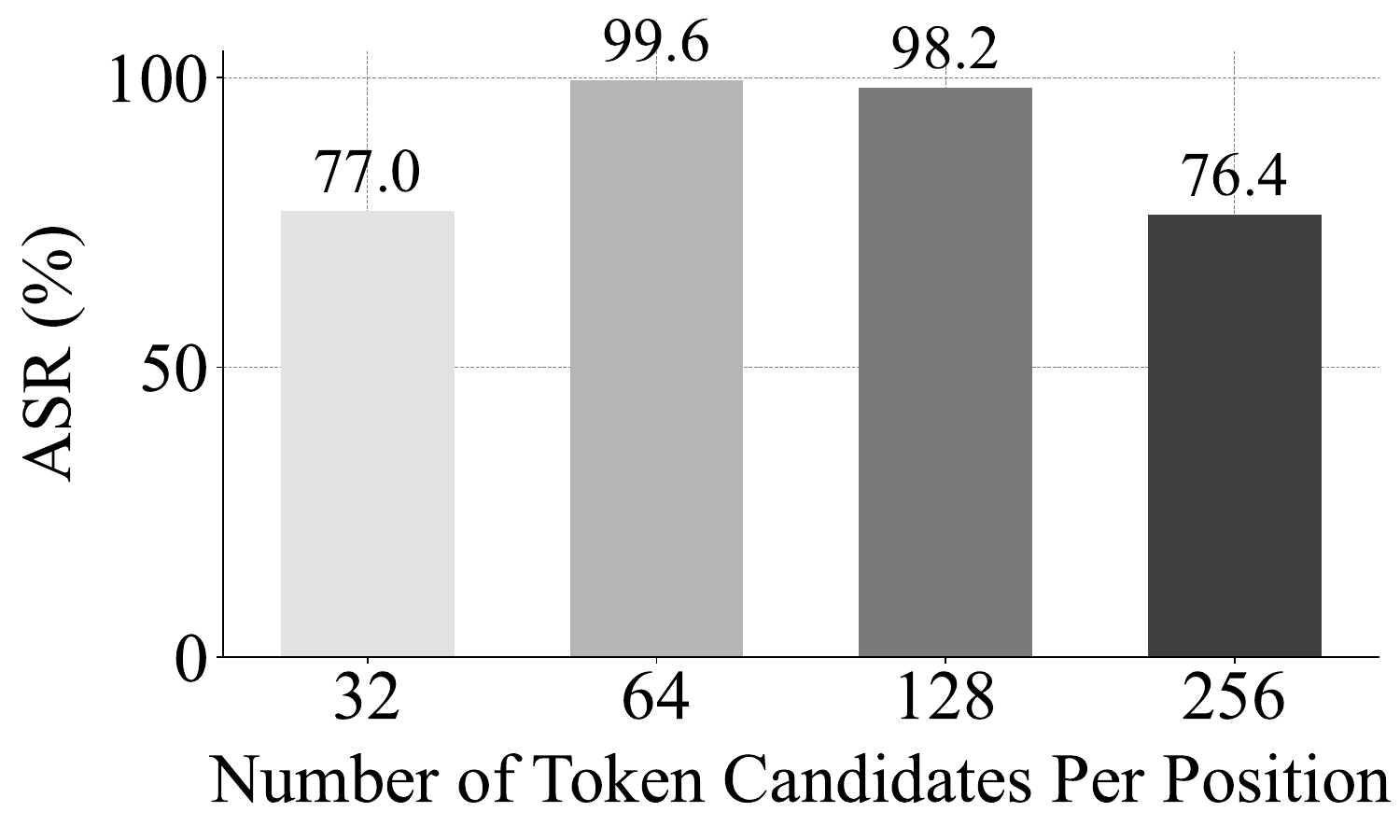}\label{fig:AS_d_tok}}
    \subfloat[]{\includegraphics[width=0.31\linewidth]{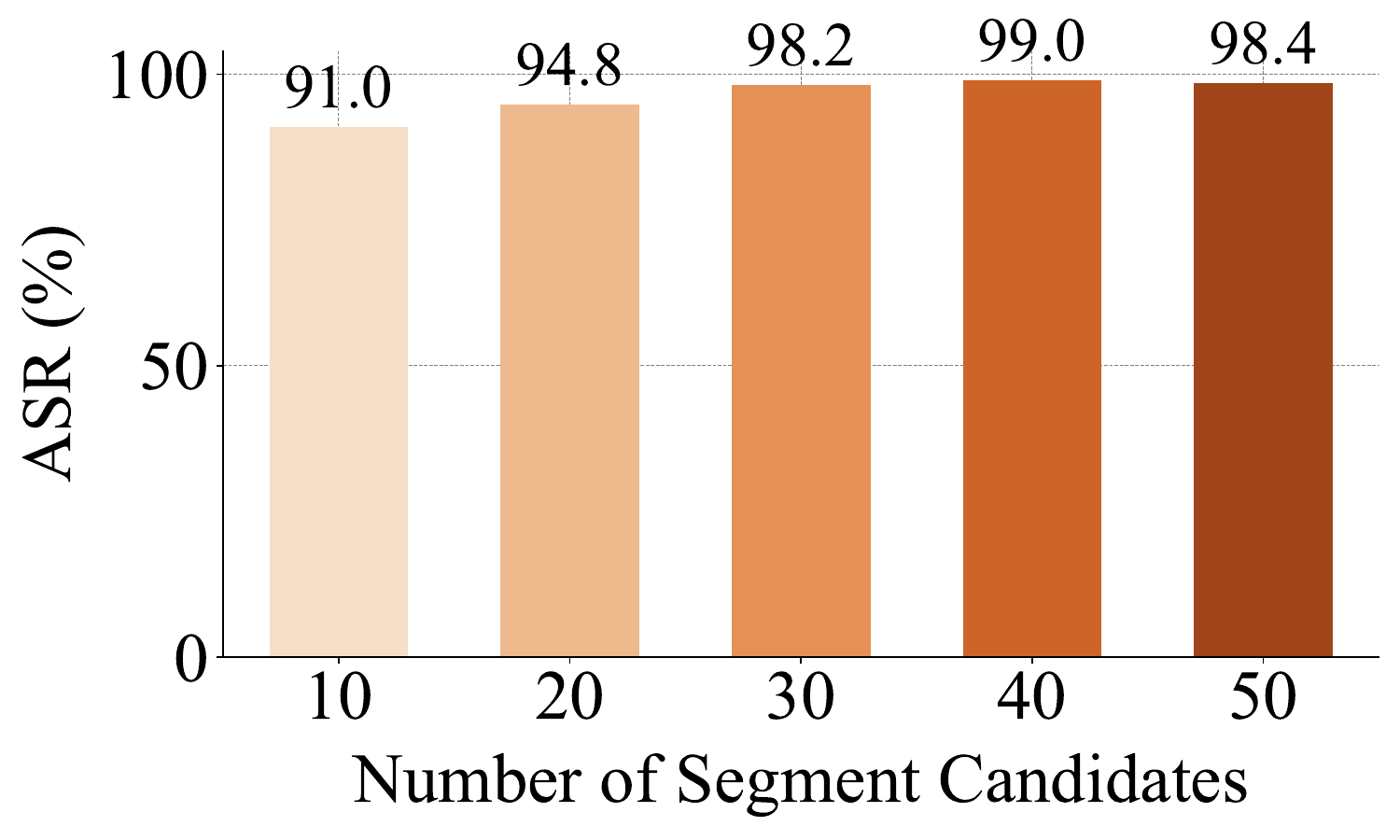}\label{fig:AS_d_seg}}
    \subfloat[]{\includegraphics[width=0.31\linewidth]{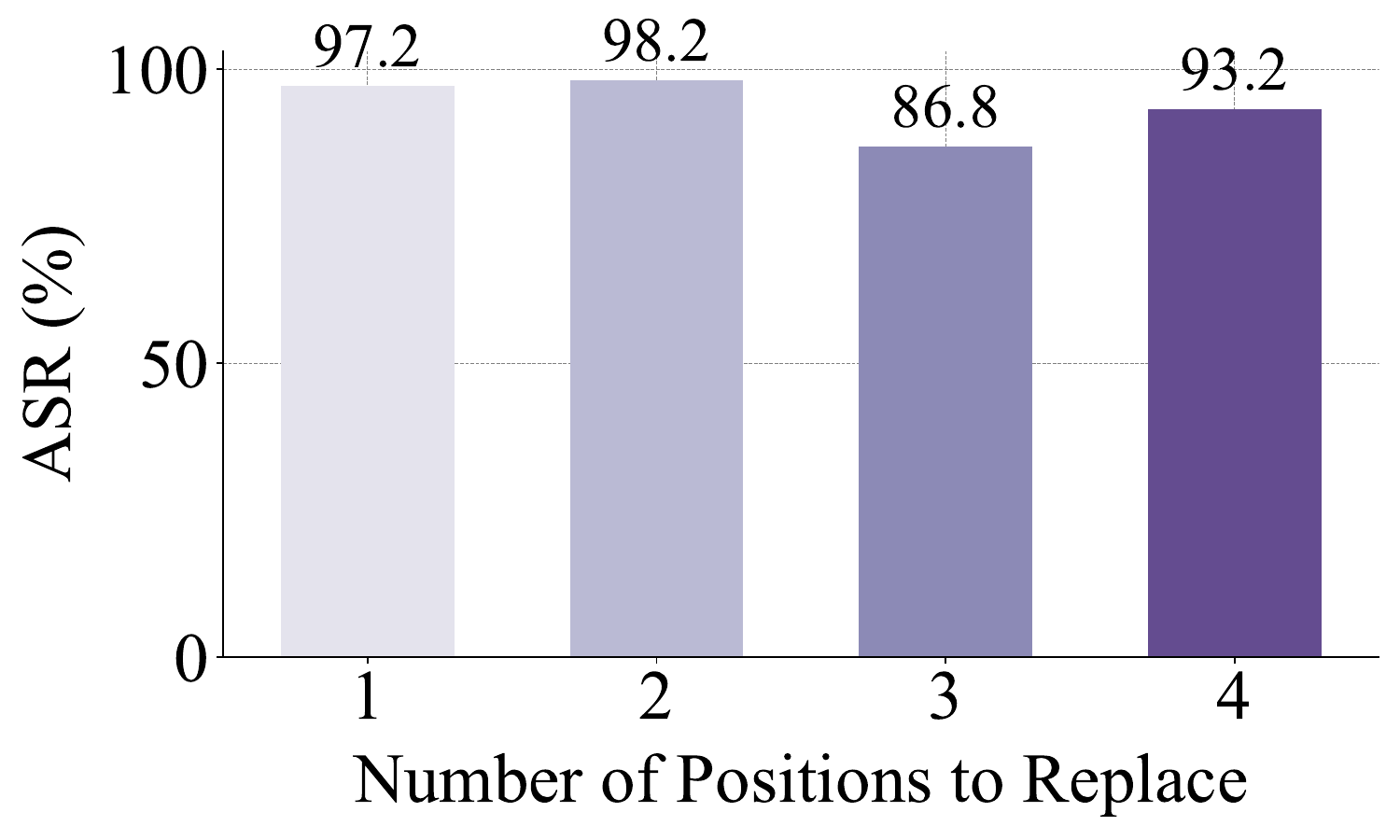}\label{fig:AS_d_rep}}
    
    \caption{Impact of (a) number of token candidates per position $d_\text{tok}$, (b) number of segment candidates $d_\text{seg}$, and (c) number of positions $d_\text{rep}$ to replace in a segment on the ASR of \method{}.}
    \label{fig:appendix_hyperparameter}
\end{figure*}

\begin{table*}[!t]
    \centering
    \caption{ASR (\%) of ObliInjection against leave-one-segment-out and segment delimiters defenses.}
    \label{tab:basic_defenses}
    \addtolength{\tabcolsep}{-2pt}

        \begin{tabular}{l cccccccc}
            \toprule
            \textbf{Defense}
                & \textbf{Llama-3-8B}
                & \textbf{Llama-3.1-8B}
                & \textbf{Mistral-7B}
                & \textbf{Qwen-2.5-7B}
                & \textbf{Falcon3-7B}
                & \textbf{Llama4-17B}
                & \textbf{Qwen3-4B}
                & \textbf{Average} \\
            \midrule

            Leave-one-segment-out
                & 99.2 & 98.9 & 99.8 & 100.0 & 99.2 & 98.4 & 99.4 & 99.3 \\

            Segment Delimiters
                & 100.0 & 91.4 & 98.2 & 93.6 & 91.6 & 99.8 & 99.2 & 96.3 \\

            \bottomrule
        \end{tabular}
     \vspace{2mm}
\end{table*}

\begin{table*}[!t]
\caption{Our prompt used to query the auxiliary LLM to generate a shadow target instruction for each dataset.}
\centering
\begin{tabular}{l p{14cm}}
\toprule
\textbf{Dataset} & \textbf{Prompt} \\
\midrule
Amazon Reviews & \textit{You are a prompt designer. Your task is to craft a clear and effective prompt for LLM to summarize Amazon product reviews. The summary should highlight key strengths, weaknesses, and the overall product highlights. Make sure the prompt is suitable for use in an automated pipeline and encourages informative, customer-friendly summaries.} \\
Multi-News & \textit{You are a prompt designer. Your task is to write a clear and effective prompt for LLM to summarize news articles. The summary should capture the main events, key facts, and overall context in a concise and neutral tone. Ensure the prompt is suitable for automated summarization and encourages factual, well-structured output.} \\
HotpotQA & \textit{You are a prompt designer. Your task is to write a clear and effective prompt for LLM to answer a question based on multiple source documents or materials. The prompt should encourage the model to reason across the materials, synthesize relevant information, and provide a coherent, well-supported answer. Ensure the prompt is suitable for multi-hop question answering tasks.}\\
\bottomrule
\end{tabular}

\label{tab:prompt_generate_ins}
\end{table*}

\begin{table*}[!t]
\caption{Prompt used to query the auxiliary LLM to generate shadow segments for the Amazon Reviews dataset.}
\centering
\begin{tabular}{l p{14cm}}
\toprule
\textbf{Dataset} & \textbf{Prompt} \\
\midrule
Amazon Reviews & 
\begin{minipage}[t]{14cm}
\itshape
You are asked to generate {100 unique customer reviews} based on the product details provided below. These reviews should simulate real user experiences with a variety of tones and formats.

{Product Metadata:}
\begin{itemize}
    \item {Category:} \emph{\{Category from metadata $\mathcal{M}^t$\}}
    \item {Name:} \emph{\{Product name from metadata $\mathcal{M}^t$\}}
    \item {Features:} \emph{\{Product features from metadata $\mathcal{M}^t$\}}
    \item {Description:} \emph{\{Product description from metadata $\mathcal{M}^t$\}}
    \item {Price:} \emph{\{Product price from metadata $\mathcal{M}^t$\}}
\end{itemize}

Each review should:
\begin{itemize}
    \item Address aspects such as quality, performance, usability, durability, and value
    \item Align with the rating in tone and sentiment
    \item Vary in style—ranging from concise comments and detailed narratives to pros/cons lists and casual formats with typos or emojis
\end{itemize}

Output Format:

\{
    ``Title": ``[Review title]",
    ``Text":``[Review text]",
    ``Rating": [Rating from 1 to 5]
\}

Example:

\{
    ``Title": ``Five Stars",
    ``Text": ``On time. Works great!",
    ``Rating": 5.0
\}

Now generate {100 reviews} that follow the above guidelines and metadata.
\end{minipage}
\\
\bottomrule
\end{tabular}

\label{tab:prompt_amazon}
\end{table*}

\begin{table*}[!]
\caption{Prompts used to query the auxiliary LLM to generate shadow segments for Multi-News and HotpotQA datasets.}
\centering
\begin{tabular}{l p{14cm}}
\toprule
\textbf{Dataset} & \textbf{Prompt} \\
\midrule
Multi-News & 
\begin{minipage}[t]{14cm}
\itshape
You are asked to write {10 synthetic news articles} based on the event summary provided below. Each article should reflect the style and voice of a different media outlet.

{News Metadata:}
\begin{itemize}
    \item {Event Summary:} \emph{\{Key facts from metadata $\mathcal{M}^t$\}}
\end{itemize}

Each article should:
\begin{itemize}
    \item Present a distinct journalistic voice (e.g., objective, opinionated, local, informal)
    \item Vary in structure, tone, and length (between 100–500 words)
    \item Rephrase and elaborate on the provided facts without copying them verbatim
    \item Read fluently and realistically, as if written by a professional journalist
\end{itemize}

Output Format:

\{
    ``Title": ``[Headline (1 to 20 words)]",
    ``Text": ``[News article (100 to 500 words)]"
\}

Example:

\{
    ``Title": ``Community Rallies After Storm Damage",
    ``Text": ``Local residents and volunteers are working together..."
\}

Now generate {10 news articles} that follow the above requirements and metadata.
\end{minipage}
\\
\addlinespace
HotpotQA & 
\begin{minipage}[t]{14cm}
\itshape
You are asked to generate a question and a list of supporting facts based on the provided question type. Write {1 natural-sounding question} followed by {10 supporting facts}, each written as if sourced from a credible publication.

{Question Metadata:}
\begin{itemize}
    \item {Question Type:} \emph{\{Type from metadata $\mathcal{M}^t$\}}
\end{itemize}

Each supporting fact should:
\begin{itemize}
    \item Be relevant, informative, and topically related to the question
    \item Mix factual and plausible (but fictional) content
    \item Appear well-grounded and realistic in style and phrasing
\end{itemize}

Output Format:

\{
    ``Question": ``[The question]",
    ``Title": ``[Informative title for the fact]",
    ``Text": ``[100 to 300 word paragraph]"
\}

Example:

\{
    ``Question": ``Are Jean Genet and Mark Sandrich both from France?",
    ``Title": ``Mark Sandrich: Hollywood Director",
    ``Text": ``Mark Sandrich was a prominent American film director..."
\}

Now generate {the question and 10 supporting facts} according to the above instructions.
\end{minipage}
\\
\bottomrule
\end{tabular}

\label{tab:prompt_other_datasets}
\end{table*}

\subsection{Additional Experiments}
\label{appendix:additional}
\myparatight{Other injected responses} To evaluate the effectiveness of \method{} for different injected responses, we conduct experiments on the Amazon Reviews dataset using three variants of $r^e$: ``The product is useless!'', ``The product is amazing!'', and ``The product is average.'' These reflect different attacker intents, such as demoting or promoting a product, while the injected task is still review summarization. During optimization, we keep the injected prompt $p^e$ consistent with the corresponding response when constructing the contaminated segment. We consider the model output $f(s^t||x^c)$ to be semantically equivalent to $r^e$ if it contains the keyword ``useless'', ``amazing'', or ``average'', respectively. As shown in Table~\ref{tab:different_injected_data}, \method{} achieves high ASR across all three cases--99.0\% for ``The product is useless!'', 97.7\% for ``The product is amazing!'', and 98.8\% for ``The product is average.'' This demonstrates that \method{} can be effectively tailored to different attacker goals with various injected responses.

\begin{table*}[!t]
    \centering
    \caption{ASR (\%) of \method{} for different injected responses across LLMs.}
    \label{tab:different_injected_data}
    \begin{tabular}{l cccccccc}
        \toprule
        \textbf{Injected response $r^e$}
            & \textbf{Llama-3-8B}
            & \textbf{Llama-3.1-8B}
            & \textbf{Mistral-7B}
            & \textbf{Qwen-2.5-7B}
            & \textbf{Falcon3-7B}
            & \textbf{Llama4-17B}
            & \textbf{Qwen3-4B}
            & \textbf{Average} \\
        \midrule
        The product is useless!
            & 99.4 & 98.0 & 99.2 & 99.8 & 98.2 & 99.8 & 98.8 & 99.0 \\
        The product is amazing! 
            & 94.0 & 99.4 & 94.0 & 100.0 & 99.4 & 99.6 & 97.6 & 97.7 \\
        The product is average.
            & 100.0 & 100.0 & 95.4 & 98.2 & 98.6 & 100.0 & 99.4 & 98.8 \\
        \bottomrule
    \end{tabular}
\end{table*}

\myparatight{Other injected tasks} The experiments above assume that the injected task is text summarization, while varying the injected responses. To evaluate the effectiveness of \method{} across a broader range of injected tasks, we conduct experiments on five types of natural language tasks, following Liu et al.~\cite{liu2024prompt}: duplicate sentence detection using the MRPC~\cite{dolan-brockett-2005-automatically} dataset, hate content detection using HSOL~\cite{hateoffensive}, natural language inference using RTE~\cite{wang2019glue}, sentiment analysis using SST2~\cite{socher-etal-2013-recursive}, and spam detection using the SMS Spam~\cite{Almeida2011SpamFiltering} dataset. For each type of task, we randomly sample one injected task $(s^e, x^e, r^e)$ to construct a contaminated segment in the form $z||p^e||z'$, where $p^e = s^e||x^e$. All other settings follow the default configuration described in Section~\ref{sec:expsetup}. During evaluation, we consider the attack successful if the model's output $f(s^t || x^c)$ exactly matches $r^e$, where $x^c$ is the contaminated data formed by randomly permuting the clean and contaminated segments.

As shown in Table~\ref{tab:different_injected_task}, \method{} demonstrates strong attack effectiveness across all five injected tasks and five LLMs, achieving average ASRs of 97.7\% on duplicate sentence detection, 97.3\% on hate detection, 95.4\% on natural language inference, 97.4\% on sentiment analysis, and 94.2\% on spam detection. Each LLM achieves an ASR above 91\%, highlighting the effectiveness of \method{} across both diverse tasks and model architectures.

\begin{table*}[!t]
    \centering
    \caption{ASR (\%) of \method{} for different injected tasks across LLMs.}
    \label{tab:different_injected_task}
    \addtolength{\tabcolsep}{-3pt}
    \begin{tabular}{l cccccccc}
        \toprule
        \textbf{Injected task}
            & \textbf{Llama-3-8B}
            & \textbf{Llama-3.1-8B}
            & \textbf{Mistral-7B}
            & \textbf{Qwen-2.5-7B}
            & \textbf{Falcon3-7B}
            & \textbf{Llama4-17B}
            & \textbf{Qwen3-4B}
            & \textbf{Average} \\
        \midrule
        Duplicate Sentence Detection
            & 94.6 & 96.8 & 96.4 & 99.2 & 97.8 & 100.0 & 98.8 & 97.7 \\
        Hate Detection
            & 100.0 & 99.0 & 91.0 & 95.8 & 99.8 & 96.0 & 99.5 & 97.3 \\
        Natural Language Inference
            & 97.0 & 91.4 & 98.0 & 94.0 & 96.8 & 91.0 & 99.6 & 95.4 \\
        Sentiment Analysis
            & 98.4 & 96.6 & 92.0 & 100.0 & 98.0 & 100.0 & 97.0 & 97.4 \\
        Spam Detection
            & 91.2 & 92.8 & 93.8 & 91.6 & 99.0 & 100.0 & 91.2 & 94.2 \\
        \bottomrule
    \end{tabular}
    \vspace{-2mm}
\end{table*}

\myparatight{Other hyperparameters of \method{}} Figure~\ref{fig:appendix_hyperparameter} shows how the ASR on Falcon3-7B varies with three hyperparameters of \method{}: the number of token candidates per position $d_\text{tok}$ in Algorithm~\ref{alg:generate_candidate_tokens_each_position}, the number of segment candidates $d_\text{seg}$ in Algorithm~\ref{alg:generate_sequence_candidate}, and the number of positions $d_\text{rep}$ to replace in a segment in Algorithm~\ref{alg:generate_sequence_candidate}. For $d_\text{tok}$,  too small or too large values lead to lower ASR. A too small $d_\text{tok}$ limits new token candidates to those with low estimated loss, which may be inaccurate due to approximation errors in the Taylor expansion. Conversely, a too large $d_\text{tok}$ introduces high-loss candidates that can mislead optimization. For $d_\text{seg}$, increasing its value generally improves  ASR, as more segment candidates increases the likelihood of identifying one that yields a higher ASR. Finally, ASR drops when $d_\text{rep}$ is too large (e.g., exceeds 2), likely because the new segment diverges too much from the original, reducing the attack's effectiveness.

\begin{table}[!t]\renewcommand{\arraystretch}{1.2}
\addtolength{\tabcolsep}{-4pt}
  \centering
  \fontsize{9}{9}\selectfont
  \caption{Important notations.} 
  \vspace{-2mm}
  \begin{tabular}{|c|c|}\hline
    \textbf{Notation} & \textbf{Description} \\ \hline \hline
    $f$ & LLM \\ \hline 
    Superscript $^t$ & Information about target task \\ \hline
    Superscript $^e$ & Information about injected task \\ \hline
    $s^t$, $x^t$, or $p^t$ & Target instruction, data, or prompt \\ \hline 
    $s^e$, $x^e$, or $p^e$ & Injected instruction, data, or prompt \\ \hline 
    $x^c$ & Contaminated data \\ \hline 
    $p^c=s^t|| x^c$ & Contaminated target prompt \\ \hline 
    $n$ & Number of data sources \\ \hline 
    Subscript $_s$ & Shadow information \\ \hline
    $s^t_s$ & Shadow target instruction \\ \hline
    $n_s$ & Shadow number of data sources \\ \hline 
    $x^t_i$ & Segment from $i$th source \\ \hline     
    $x$ & Contaminated segment \\ \hline 
    $x_j$ & $j$th token of $x$ \\ \hline 
    $x_s^{(i)}$ & The $i$th shadow segment for the target task  \\ \hline 
    $\mathcal{X}_s$ & Set of shadow segments for the target task \\ \hline 
    $d_{\text{iter}}$ & Number of iterations \\ \hline 
    $d_{\text{per}}$ & \makecell{Number of permutations to approximate \\the order-oblivious cross-entropy loss} \\ \hline 
    $d_{\text{tok}}$ & Number of candidates per token \\ \hline 
    $d_{\text{seg}}$ & Number of segment candidates \\ \hline 
    $d_{\text{buf}}$ & Buffer size \\ \hline 
    $d_{\text{rep}}$ & Number of positions to replace in a segment \\ \hline 
  \end{tabular}
  \label{tab:notation}
\end{table}

\end{document}